\documentclass[amssymb,twocolumn,aps]{revtex4}
\usepackage{graphics,color,graphicx,amsmath}
\usepackage{subcaption}
\usepackage{natbib}
\usepackage{verbatim}
\usepackage{graphicx}
\usepackage[above,below]{placeins}
\usepackage{float}
\usepackage{tabularx}
\usepackage{times}
\usepackage{blindtext}
\usepackage{url}
\usepackage{rotating}

\begin{document}

\title{Who and what gets recognized in peer recognition

%Probing the nature of gender bias in student nominations of strong physics peers
}

\author{Meagan Sundstrom,~\footnote[1]{mas899@cornell.edu}$^{1}$ L. N. Simpfendoerfer,$^1$ Annie Tan,$^1$ Ashley B. Heim,$^2$ and N. G. Holmes~\footnote[2]{Corresponding author. ngholmes@cornell.edu}$^1$}
\affiliation{$^1$Laboratory of Atomic and Solid State Physics, Cornell University, Ithaca, New York 14853, USA\\
$^2$Department of Ecology and Evolutionary Biology, Cornell University, Ithaca, New York 14853, USA}

\date{\today}

\begin{abstract}

Previous work has identified that recognition from others is an important predictor of students' participation, persistence, and career intentions in physics. However, research has also found a gender bias in peer recognition in which student nominations of strong peers in their physics course disproportionately favor men over women. In this study, we draw on methods from social network analysis and find a consistent gender bias in which men disproportionately under-nominate women as strong in their physics course in two offerings of both a lecture course (for science and engineering, but not physics, majors) and a distinct lab course (for science, engineering, and physics majors). We also find in one offering of the lecture course that women disproportionately under-nominate men, contrary to what previous research would predict. We expand on prior work by also probing two data sources related to who and what gets recognized in peer recognition: students' interactions with their peers (who gets recognized) and students' written explanations of their nominations of strong peers (what gets recognized). %We find that students determine who gets recognized in two different ways, each with a similar frequency: (1) interaction-based recognition, where students nominate peers with whom they directly interact, and (2) observation-based recognition, where students indirectly observe peers with whom they do not interact. 
Results suggest that the nature of the observed gender bias in peer recognition varies between the instructional contexts of lecture and lab. In the lecture course, the gender bias is related to who gets recognized: both men and women disproportionately over-nominate their interaction ties to students of their same gender as strong in the course. In the lab course, the gender bias is also related to what gets recognized: men nominate men more than women because of skills related to interactions, such as being helpful. These findings illuminate the different ways in which students form perceptions of their peers and add nuance to our understanding of the nature of gender bias in peer recognition.
\end{abstract}

\maketitle

\section{Introduction}

Gaining recognition as a physicist is important for students' participation and persistence in their physics course~\cite{hyater2018critical,lock2013physics,carlone2007understanding,hazari2010connecting,hazari2018towards,hazari2017importance,kalender2019gendered,boe2023cleverness}. Recognition is particularly important for the participation of historically underrepresented groups in physics, such as women~\cite{carlone2007understanding,kalender2019gendered,avraamidou2022identities,li2023womenperceivedrecognition}. However, research has found that men both perceive higher recognition from others~\cite{lock2013physics,kalender2019gendered,hazari2013science,bottomley2022relationship} and receive more recognition from their physics peers~\cite{bloodhart2020,sundstrom2022perceptions} than women. To better understand these effects, we investigated the nature of student recognition of strong peers with a focus on the gender bias in such peer recognition. Specifically, we probe two questions: whether and how gender bias in students' nominations is related to patterns of peer interactions (\textit{who} gets recognized), and whether and how gender bias in students' nominations is related to their written explanations of these nominations (\textit{what} gets recognized). Throughout the paper, we use the phrase \textit{gender bias} to refer to a distinguishable difference between the amount of peer recognition received by men (women) and the amount of peer recognition we would expect men (women) to receive if recognition were distributed equitably across men and women.

\subsection{Recognition in physics courses}

A student's sense of physics identity -- the degree to which they believe they are a ``physics person"~\cite{gee2000chapter} -- has been shown to predict their participation, persistence, and career intentions in physics~\cite{hazari2010connecting,kalender2019gendered,lock2013physics}. Researchers have modeled physics identity as containing three dimensions: performance and competence, interest, and recognition~\cite{carlone2007understanding,hazari2010connecting}. Previous studies demonstrate that \textit{recognition} is the most important of these four dimensions in relating to and predicting student outcomes~\cite{hyater2018critical,lock2013physics,carlone2007understanding,hazari2010connecting,hazari2018towards,hazari2017importance,kalender2019gendered,boe2023cleverness}. Recognition is the extent to which meaningful others (e.g., peers, teachers, and family) perceive an individual as a physics person. When a student receives more recognition from others, they are more likely to see themselves as a physics person and therefore develop a stronger physics identity~\cite{hazari2018towards, kalender2019female}.

Recognition from others, however, is often shaped by sociohistorical norms and stereotypes, such as those that position men as more suitable to the field of physics than women~\cite{hasse2002gender,danielsson2012exploring,gonsalves2016masculinities,kessels2006goes,makarova2015trapped,makarova2019,tate2005does,ceglie2011underrepresentation,moss2012,eddy2014,carlone2007understanding,kalender2019gendered,avraamidou2022identities}. Perhaps as a result of such stereotypes, a handful of research studies demonstrate that men report higher perceived recognition (the extent to which they feel recognized by others) in their physics classes than women~\cite{lock2013physics,kalender2019gendered,hazari2013science,bottomley2022relationship}. This difference may put men in a better position than women to develop their physics identity, contributing to the underrepresentation of women in physics documented in, for example, Refs.~\cite{sax2016,johnson2020women,porter2019women}.

%minoritized groups such as women often receive less recognition than their peers~\cite{grunspan2016,bloodhart2020,sundstrom2022perceptions}.  %Gonsalves and colleges~\cite{gonsalves2016masculinities}, for example, conducted an ethnographic study of an experimental research group. Several PhD students in the group indicated that performing physics experiments well is associated with masculinity and that men have a natural ability to do physics. The researchers also observed a division of labor when the senior graduate students assigned tasks to the more junior members of the group: the men were to deconstruct the equipment while the women were to organize tools and clean the lab space. Perhaps as a result of such stereotypes, a handful of research studies demonstrate that men report lower perceived recognition (the extent to which they feel recognized by others) in their physics classes than their female peers~\cite{lock2013physics,kalender2019gendered,hazari2013science}. 

\subsection{Gender bias in peer recognition}

Other studies have probed ``actual" peer recognition, rather than perceived recognition, by asking students to nominate peers they believe are strong in their science course~\cite{grunspan2016,salehi2019,bloodhart2020,sundstrom2022perceptions}. We use the term ``actual" recognition to mean a measure of how much others recognize an individual as a physics person, though the others may not have indicated that recognition to the individual, necessarily. These studies largely draw on quantitative methods of social network analysis to determine the extent to which a gender bias exists in students' nominations of strong peers~\cite{grunspan2016,salehi2019,sundstrom2022perceptions}, finding mixed results. Grunspan and colleagues~\cite{grunspan2016}, for example, examined three offerings of an introductory biology course (the second in the course sequence) for first-year students. They observed that men disproportionately under-nominated women, while women proportionately nominated men and women, as strong in the course material in all three offerings. Bloodhart and colleagues~\cite{bloodhart2020} observed a similar gender bias in peer recognition across many introductory physics courses for first-year students, but found that women also disproportionately under-nominated women as strong in the course material. In the same study, researchers found that men proportionately nominated both men and women, but women disproportionately over-nominated other women, as strong in introductory life sciences courses for first-year students. Salehi and colleagues~\cite{salehi2019}, however, found no gender bias in either men's or women's nominations of strong peers across two offerings of a mechanical engineering course taken by second and third-year students. 

Our previous work~\cite{sundstrom2022perceptions} examined peer recognition in three different remote physics courses and added nuance to these studies. We observed a gender bias in peer recognition favoring men (in which men disproportionately under-nominated women, but women proportionately nomination men and women) in two introductory physics courses aimed at first-year students, but no gender bias favoring men in an introductory physics course comprised mostly of second-year students (though women disproportionately over-nominated other women in this course). Comparing across all four studies~\cite{grunspan2016,salehi2019,bloodhart2020,sundstrom2022perceptions}, the presence or absence of a gender bias in peer recognition seems to vary by course level more than any other aspect of the instructional context: researchers find a gender bias in peer recognition in science courses for first-year, but not beyond-first-year, students. Gender bias in peer recognition also seems related to student outspokenness (i.e., verbally participating in class). In the two of these four studies that measure outspokenness, a gender bias in students' nominations of strong peers (favoring men) is present when there is also a gender disparity in who is outspoken (i.e., when men participate more than women), and there is no gender bias in peer recognition when there is no gender disparity in who is outspoken~\cite{grunspan2016,sundstrom2022perceptions}.

Patterns of peer recognition also seem to vary by instructional context, with our previous study of physics courses~\cite{sundstrom2022perceptions} finding more evidence of gender bias in peer recognition in the context of lecture material than lab material. This result may be attributable to these two instructional contexts covering distinct content and aiming to develop different sets of skills~\cite{Phys21,kozminski2014aapt,holmes2018introductory,Smith2021,smith2020direct,holmes2015teaching}. Indeed, research has shown that students believe lecture skills include knowing mathematics, while lab skills involve handling equipment and using technical skills~\cite{gonsalves2016masculinities,danielsson2012exploring}. However, the difference in peer recognition across instructional contexts may also be attributable to pedagogy: lectures typically contain many students who focus on the instructor and labs typically contain a small number of students who collaborate on tasks. This variation in how much visibility students have in front of their peers may impact patterns of peer recognition, especially during remote instruction as in our prior work~\cite{sundstrom2022perceptions}. In the current study, therefore, we determine the extent to which gender bias in peer recognition exists when lecture and lab material are taught with a similar pedagogical style and in-person. Specifically, we analyze physics courses for first-year students where the lab and lecture material comprise distinct courses (i.e., students co-enroll in one lab course and one lecture course and receive a separate grade in each course), and both of these courses contain a lecture session where all students focus on the instructor and a small-group session where students solve problems or conduct experiments with their peers.

\subsection{What is the nature of this gender bias?}

While the quantitative studies mentioned above importantly determine the extent to which a gender bias in peer recognition exists, they do not probe the nature of this gender bias. Toward this end, separate threads of research have started to unpack two possible mechanisms underlying recognition: peer interactions (i.e., students learn about their peers' skills during interactions with these peers) and students' reasons for recognizing others as strong in their physics course (i.e., students recognize their strong physics peers for different skills). Of course, there may also be other explanations for the gender bias in peer recognition, such as sociohistorical gender stereotypes alone, but, in this study, we seek to build on the existing research threads.

\subsubsection{Who gets recognized: Peer interactions and indirect observations of peers}

Previous work suggests that one mechanism through which recognition forms is interactions with others~\cite{gee2000chapter,quan2022trajectory,alaee2022}. In their original conception of the identity framework, Gee states that ``the modern need for recognition places a particular importance on discourse and dialogue...Individuals must win recognition for them through exchange with others"~\cite[p. 112-113]{gee2000chapter}. We interpret this to mean that an individual demonstrates their knowledge, skills, and personality traits through conversations with others, who then form perceptions of that individual as a certain kind of person. %Within the context of a physics classroom, for example, students might ``win" recognition from their peers as a physics person if they correctly solve problems in front of them. 
One study, for example, conducted interviews with undergraduate students to understand their experiences in a remote summer research program~\cite{alaee2022}. The authors found that students' research group members and advisors started to recognize the students as physicists and researchers during interactions with one another: ``Other recognition was supported by conversations between the mentee and other group members"~\cite[p. 10]{alaee2022}.

Similar work describes peer interactions as a way for students to determine who of their peers is strong in physics. % and understand the skills associated with being “good” at physics -- what it meant to be recognized as a strong physics student. 
In one study, researchers performed a longitudinal case study of a woman in physics named Cassidy~\cite{quan2022trajectory}. At the beginning of her undergraduate physics studies, Cassidy recognized her more senior peer tutor, one of the only other physics students with whom she interacted, as a ``smart" physics student because they showed her an unnecessarily complicated solution to a physics problem. About a year later, however, Cassidy became a more outgoing member of the physics community who frequently collaborated with peers on assignments. These peer interactions facilitated Cassidy’s understanding of the ``multiplicity of ways to be `good' at physics,"~\cite[p. 12]{quan2022trajectory} such as bringing in different areas of expertise to a peer collaboration. She then recognized many of her peers as being strong physics students, rather than only her peer tutor. This and other studies~\cite{gee2000chapter,alaee2022}, therefore, suggest that interactions are likely a mechanism for forming peer recognition: interactions facilitate students' understanding of their peers' knowledge and skill sets which informs who gets recognized. This mechanism of forming peer recognition may also relate to gender bias in peer recognition because previous work has found that students tend to interact with peers of their same gender~\cite{dokuka2020academic, sundstrom2022interactions}.

Peer interactions, however, are not the only means through which students determine who gets recognized. Grunspan and colleagues~\cite{grunspan2016}, for example, demonstrate that outspokenness -- frequent verbal participation in front of many others -- also relates to which students receive peer recognition. They found that students who actively participated in lecture tended to receive more nominations from peers as strong in their biology course despite these students never directly interacting with one another. %Similarly, in physics lab courses where several lab groups work in close proximity to one another, students can likely observe students in other lab groups and may recognize them as strong in physics without collaborating with them.
In addition to direct interactions with peers, therefore, students may determine who they recognize as a strong peer by indirectly observing their peers.

In the current study, we examine the relationship between peer interactions, indirect observations of peers, and peer recognition by quantitatively comparing students' self-reported peer interactions to their nominations of strong peers. We also compare this relationship across men's versus women's nominations to determine whether patterns of interactions relate to the nature of gender bias in peer recognition (we could not measure whether patterns of indirect observations relate to the nature of gender bias in peer recognition, see Sec.~\ref{sec:intanalysis}).

\subsubsection{What gets recognized: Skill sets associated with being a physicist}

Other studies have explicitly probed the knowledge, skill sets, and traits for which students recognize strong peers in their physics or other science courses~\cite{tonso2006student,danielsson2012exploring,fields2013picking,due2014competent,gonsalves2014physics,gonsalves2014persistent,gonsalves2016masculinities,irvingsayre2015,DoucetteHermione,doucettegoodlabpartner,cooper2018perceives,stump2022perc}. Doucette and colleagues, for example, asked undergraduate physics students to describe their ideal lab partner~\cite{doucettegoodlabpartner}. The authors identified 13 characteristics from the responses that students recognized in a ``good lab partner,'' including knowledgeable, hardworking, communicative, helpful, and efficient. In another study, Irving and Sayre interviewed upper-level physics students and asked them what they think it means to be a physicist~\cite{irvingsayre2015}. The participants noted a wide array of skills or traits that they recognized in a physicist, including intuition for learning physics, interest in physics, solving physics problems, designing experiments, and collecting and interpreting experimental data. Some studies also relate the identified skills to gender. Danielsson~\cite{danielsson2012exploring}, for instance, found that undergraduate students associate natural ability, tinkering with lab equipment, and mathematical competence with men and diligence and note-taking with women. 

In the current study, we expand on this body of work by (i) collecting and analyzing a large number of student explanations of their nominations of strong peers and (ii) comparing the frequencies of explanations written by men and women when nominating men versus women to determine whether and how the explanations relate to gender bias in the nominations. %We also explore these explanations across distinct instructional lecture and lab contexts.

%students' actual nominations of their strong physics peers (i.e., gender bias).

%, however most of these studies do not examine a large number of students. These studies also do not relate their qualitative data to relevant quantitative data, such as students' actual nominations of their strong physics peers. 

\subsection{Current study}

%As mentioned, gaining recognition from peers is important for students' development of their physics identity. However, peer recognition is often impacted by sociohistorical stereotypes that typically position men as more suitable to the field than women.

In summary, research has demonstrated that whether a gender bias exists in peer recognition varies across courses and instructional contexts. Prior work also suggests that peer interactions and differences in what skills sets are associated with being strong in physics might help to explain the nature of this gender bias. To probe these two possible mechanisms underlying who and what gets recognized in peer recognition, we conducted a mixed-methods study of in-person physics courses to answer the following research questions: 
\begin{enumerate}
    \item To what extent does a gender bias exist in students' recognition of strong peers within distinct lab and lecture courses?
    \item \textbf{Who gets recognized:} In distinct lab and lecture courses, how (if at all) is gender bias in peer recognition related to patterns of peer interactions?
    \item \textbf{What gets recognized:} In distinct lab and lecture courses, how (if at all) is gender bias in peer recognition related to the skill sets students associate with being strong in physics?
\end{enumerate}

We collected students' nominations of strong peers, explanations for these nominations, and self-reported interactions with peers in two offerings of distinct introductory lab and lecture physics courses for first-year science and engineering students at Cornell University. Similar to prior research examining introductory biology and physics courses for first-year students~\cite{grunspan2016,bloodhart2020,sundstrom2022perceptions}, we find a gender bias in peer recognition in which men disproportionately under-nominate women compared to men in all analyzed courses. We also find that women disproportionately under-nominate men in one offering of the lecture course. Comparing the nominations of strong peers to peer interactions, we observe in most cases that the overall gender bias in peer recognition is related to gender bias in interaction-based recognition, where students disproportionately over-nominate their interaction ties to peers of their same gender. Finally, we find a difference in students' written explanations in the lab course, where men nominate men more than women because of the ways they interacted, such as being helpful, but not in the lecture course, where men and women nominate men and women for similar skill sets. %These results illuminate the ways in which students form perceptions of their peers and the nature of gender bias in peer recognition observed across different physics courses and contexts.

\begin{table*}[t]
\caption{\label{tab:demographics}%
Summary of survey response rates and self-reported student demographics for the four courses we analyzed. All analyzed students in the lecture course are also in the lab course of the corresponding semester. Percentages are relative to the number of students included in the analysis unless specified otherwise. We grouped race or ethnicity by underrepresented racial minority (URM) status, where non-URM students are those solely identifying as White and/or Asian or Asian American and URM students are those identifying as at least one of any other race or ethnicity (including American Indian or Alaska Native, Black or African American, Hispanic or Latinx, and Native Hawaiian or other Pacific Islander). We denote students' gender and race or ethnicity as ``unknown'' if they preferred not to disclose this information on the survey or if they did not complete the survey.
}
\begin{ruledtabular}
\setlength{\extrarowheight}{1pt}
\begin{tabular}{lcccc}
 & \multicolumn{2}{c}{Lab course} & \multicolumn{2}{c}{Lecture course}\\
  \cline{2-3}
  \cline{4-5}
 & Fall  & Spring  & Fall  & Spring \\
\hline
Survey response rate (\% of total enrolled) & 95\% & 98\% & 99\% & 95\% \\
Students in analysis  & 387 & 646  & 237  & 513  \\
Gender \\
\hspace{3mm}Men & 200 (51.7\%) & 293 (45.4\%)& 111 (46.9\%) & 222 (43.3\%) \\
\hspace{3mm}Women & 153 (39.5\%)  & 302 (46.7\%) & 106 (44.7\%) & 256 (49.9\%) \\
\hspace{3mm}Non-binary & 4 (1.0\%) & 2 (0.3\%) & 2 (0.8\%) & 2 (0.4\%) \\
\hspace{3mm}Unknown & 30 (7.8\%) & 49 (7.6\%) & 18 (7.6\%)   & 33 (6.4\%) \\

Race or ethnicity\\
\hspace{3mm}Non-URM & 291 (75.2\%)  & 390 (60.4\%) & 174 (73.4\%) & 307 (59.8\%) \\
\hspace{3mm}URM & 59 (15.2\%)  & 164 (25.4\%) & 39 (16.5\%) & 140 (27.3\%)\\
\hspace{3mm}Unknown & 37 (9.6\%) & 92 (14.2\%) & 24 (10.1\%) & 66 (12.9\%)\\
 
Major\\
\hspace{3mm}Physics or Engineering Physics & 57 (14.7\%)  & 51 (7.9\%) & 16 (6.8\%) & 19 (3.7\%)\\
\hspace{3mm}Engineering & 260 (67.2\%) & 475 (73.5\%)  & 178 (75.1\%)  & 403 (78.5\%) \\
\hspace{3mm}Other  & 37 (9.6\%) & 35 (5.4\%) & 20 (8.4\%) & 28 (5.5\%)\\
\hspace{3mm}Unknown  & 33 (8.5\%)  & 85 (13.2\%)  & 23 (9.7\%) & 63 (12.3\%)\\

Year \\
\hspace{3mm}First-year & 319 (82.5\%)  & 594 (92.0\%) & 189 (79.8\%) & 480 (93.6\%)\\
\hspace{3mm}Second-year  & 52 (13.4\%) & 11 (1.7\%) & 37 (15.6\%)   & 9 (1.7\%) \\
\hspace{3mm}Other or unknown & 16 (4.1\%) & 41 (6.3\%) & 11 (4.6\%)  & 24 (4.7\%)\\
\end{tabular}
\end{ruledtabular}
\end{table*}

\section{Methods}

In this section, we describe the instructional context of our study and then discuss our data collection and analysis methods.

\subsection{Instructional context}

The data come from two in-person offerings (fall and spring) of two distinct introductory physics courses (summarized in Table~\ref{tab:demographics}), one \textit{lab course} and one \textit{lecture course}, at Cornell University -- a large, private, PhD-granting institution in the northeastern United States with a Carnegie classification of very high research activity. 

The \textit{lab course} focused on developing experimental skills rather than reinforcing physics concepts~\cite[see, e.g.,][]{holmes2018introductory,Smith2021,smith2020direct,holmes2015teaching} and covered topics in both mechanics and electromagnetism. For the lab course, students attended one 50 minute lecture session (instructed by a faculty member of the physics department) and one 2 hour lab session (instructed by a graduate teaching assistant and often a supporting undergraduate teaching assistant) each week. The lecture sessions of the lab course included active learning pedagogies, such as students answering poll questions in small groups. The course was split into two lecture sections per semester, each with 200-300 students in a large stadium-seating lecture hall. During the lab sessions, which contained 20-25 students each, students conducted open-ended investigations in small groups of two to four and each group submitted lab notes at the end of every session to be graded. Lab groups were formed by the teaching assistants based on student preferences from a group-forming survey and remained the same for the whole semester. In forming the groups, the teaching assistants were advised to avoid lab groups containing an isolated woman. Outside of class, students completed individual lab homework assignments using Jupyter Notebook each week~\cite{jupyternotebookshw}. There were also multiple office hours per week where students could receive individual help on course content from graduate and undergraduate teaching assistants or the main instructor.

Most students in the lab course were simultaneously enrolled in one of two calculus-based mechanics \textit{lecture courses}: one intended for physics majors (the ``physics majors" course) and one intended for engineering and other science majors (the ``non-majors" course). In this paper, we only analyze the non-majors lecture course (200-500 students) because the physics majors course only contained 30 to 50 students. Students in this lecture course attended three 50 minute lecture sessions (instructed by a faculty member of the physics department) and two 50 minute discussion sessions (instructed by a graduate teaching assistant and often a supporting undergraduate teaching assistant) each week. This course used active learning pedagogies including a ``flipped classroom" model, such that students read relevant sections of the textbook and took a reading quiz before attending lecture. During lecture sessions, which contained half of the enrolled students at a time (there were two lecture sections per semester) and took place in a large stadium-seating lecture hall, students answered conceptual poll questions in small groups. The course also made extensive use of interactive lecture demonstrations. In the discussion sessions, which contained about 20 students each, students completed physics problems in small groups of two to four but this work was not submitted for a grade. Discussion groups were not formed by the teaching assistants, rather students formed their own groups. Students typically worked with the same discussion group every week. Outside of class, students completed individual homework assignments (problem sets) each week. There were multiple office hours per week where students typically worked together on the homework assignments with the help of graduate and undergraduate teaching assistants.

\begin{figure}[t]
    \centering
    \includegraphics[width=3in,trim={2cm 0 2cm 0}]{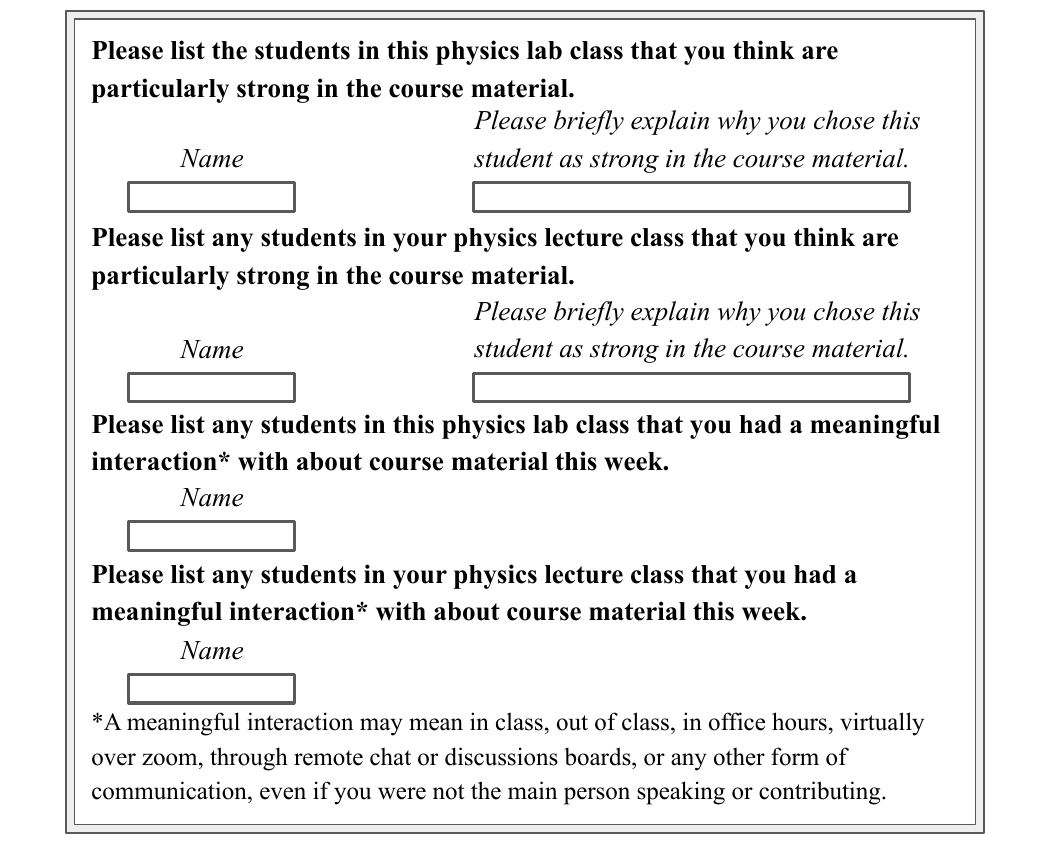}
    \caption{Online survey prompts analyzed in this study. Students typed their responses into the text boxes and could enter up to 15 peers' names for each prompt. Students were given access to the course roster.}
    \label{fig:survey}
\end{figure}

In this study, all analyzed students in the lecture course were co-enrolled in the lab course. Therefore, it was possible for students to be surrounded by some of the same peers in both courses: the lecture and lab sessions of the lab course and the lecture and discussion sections of the lecture course. Between 20\% and 40\% of students in the lab course (depending on the semester), however, were not co-enrolled in the lecture course we analyze.

%the lecture sections of each course. Lab and discussion sections contained a mix of students from the two corresponding lecture sections (i.e., lab sections contained students from both lecture sections of the lab course and discussion sections contained students from both lecture sections of the lecture course). Thus, it was also possible that students were surrounded by lecture peers (from both the lab and the lecture course) in these small sections.

\subsection{Data collection}

We administered an online network survey as part of a homework assignment in the lab course in the middle of the 15-week semester (see Fig.~\ref{fig:survey}). On the survey, we distinguished peer recognition in the lab and lecture courses because our prior work identified that patterns of peer recognition varied between these instructional contexts~\cite{sundstrom2022perceptions}. Specifically, we asked students to nominate peers in each course who they believed were knowledgeable about the course material~\cite{grunspan2016,salehi2019,bloodhart2020,sundstrom2022perceptions} as a measure of their recognition of strong peers. We also asked students to describe why they nominated their peers. 

A second set of questions asked students to self-report peers with whom they had meaningful interactions about the instructional material in each course~\cite{zwolak2018educational,traxler2020network,dou2019practitioner,commeford2021characterizing,sundstrom2022interactions}. As in prior work, ``students self-identified what counted as a meaningful interaction"~\cite[p. 6]{commeford2021characterizing}. We asked students about whom they interacted with ``this week" to capture interactions that students were consistently having with their peers throughout the semester, while reducing the possibility of recall bias (e.g., by asking them to recall all peers with whom they have interacted throughout the semester). This phrasing may have captured a few one-off interactions that only occurred the week of the survey, however these likely represent a small fraction of the reported interactions.

Each question was in an open response format, where students entered each peer's name in a separate text box and the associated explanation for each peer also in a separate text box. Students could enter up to 15 peers' names for each prompt, though no student provided the maximum number of names for any prompt. Students were also given access to the course rosters to facilitate their remembering and spelling of peers' names. Students could nominate anyone in their course; for example, they were not restricted to naming peers in their specific lecture section.

At least 95\% of enrolled students in each course responded to the survey (see Table \ref{tab:demographics}). Students occasionally misspelled peers' names and/or reported just a first or a last name. In these cases, the first author manually processed the text to match the names in the survey responses to the course roster when possible. We could not match students if the respondent provided only a first (or last) name and multiple students in the course had that first (or last) name and so these responses were subsequently dropped from the data set. In each course, we were able to match at least 90\% of the nominations to strong peers and self-reported interactions to the course roster. 

Our analysis included all students who responded to the survey and/or were listed by at least one peer on a given survey prompt. Our analysis also included only the nominations and self-reported interactions made by students who consented to participate in research (more than 95\% of survey responders). If a consenting student wrote the name of a non-consenting student, we included the survey response, but removed all information (e.g., demographics) about the non-consenting student. We were able to apply social network analysis methods to our data because both the survey response rate and the name matching rate (from the raw survey responses to the course roster) were at least 90\% and very few ($<$2\%) non-consenting students were removed from analysis, and network methods are reliable for data sets with less than 30\% missing data~\cite{smith2013structural}.

We also collected students' self-reported gender, race or ethnicity, intended major, and academic year on the survey (see Table~\ref{tab:demographics}). Most students in the data set intended to major in engineering and the majority were in their first academic year. Each offering of the lab and lecture course contained roughly equal proportions of men and women. We grouped race or ethnicity by underrepresented racial minority (URM) status, where non-URM students are those solely identifying as White and/or Asian or Asian American and URM students are those identifying as at least one of any other race or ethnicity (including American Indian or Alaska Native, Black or African American, Hispanic or Latinx, and Native Hawaiian or other Pacific Islander). The majority of students ($>$60\%) in each course were non-URM. Because the role of race or ethnicity was not part of our research questions, this categorization provided a limited ability to control for possible effects of race or ethnicity in our evaluation of the role of gender. We acknowledge, however, the limitations of this categorization~\cite{shafer2021} and encourage future work to probe this variable explicitly.

\begin{figure}[t]
    \centering
    \includegraphics[width=3in,trim={2cm 0.5cm 2cm 0.5cm}]{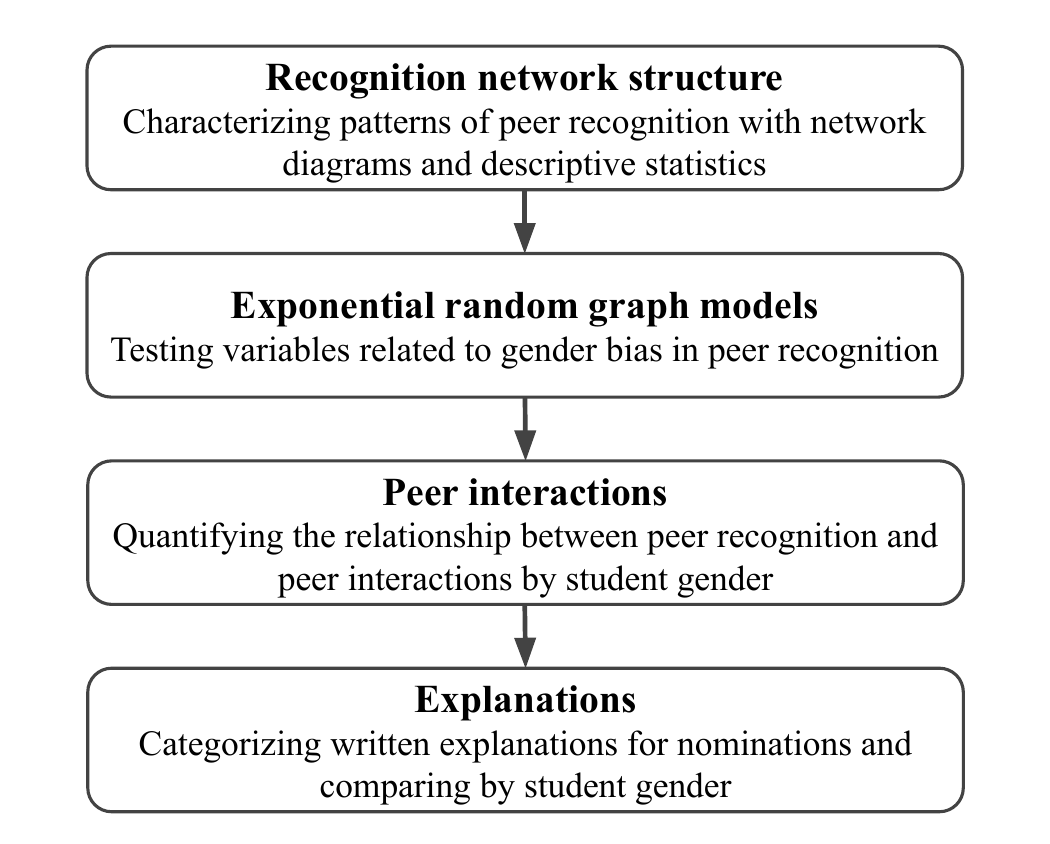}
    \caption{Flowchart depicting our stages of data analysis.}
    \label{fig:methods}
\end{figure}

At the end of the semester, we collected students' discussion and lab section enrollment, lab groups, and final grades in each course.

\subsection{Data analysis}

We conducted our data analysis in four stages (summarized in Fig.~\ref{fig:methods}), largely drawing on methods of social network analysis~\cite{grunspan2014,dou2019practitioner,brewe2018guide}.

\subsubsection{Recognition network structure}

We first analyzed the four recognition networks, one for each offering (fall and spring) of each course (lab and lecture), using student responses to the first two questions on the survey (Fig.~\ref{fig:survey}). %Despite each course having more than one lecture section, we created one network for all students in each course because students could interact with peers outside of their lecture section in the smaller sections (i.e., lab and discussion) or outside of class.
Similar to prior work~\cite{grunspan2016,salehi2019,sundstrom2022perceptions}, we converted the nominations of strong peers into directed networks (see Fig.~\ref{fig:gendersociograms}) to identify broad patterns of peer recognition. \textit{Nodes} in the network represented students and \textit{edges} (or \textit{ties}) in the network represented all nominations made between students (including direction, from the nominator to the nominee). We distinguish non-binary students from men, women, and students of unknown gender in the network diagrams (Fig.~\ref{fig:gendersociograms}) to visualize these students' positions in each network. However, non-binary students are not distinguished in the remainder of the analysis because they make up 1\% or less of the student population in each course (Table~\ref{tab:demographics}).

To characterize the structures of the observed networks, we calculated three network-level statistics -- \textit{density}, \textit{indegree centralization}, and \textit{transitivity} -- for each network. Density is the number of edges in the network that we observed as a fraction of the number of possible edges in the network. Indegree centralization measures the extent to which the nominations are concentrated around a single student or a small subset of students. This measure is calculated as the sum of differences in indegree (number of received nominations) between the node with the highest indegree (receiving the most nominations) and every other node in the network, divided by the maximum possible sum of differences of indegree for all nodes. Higher indegree centralization (i.e., closer to one) indicates higher concentration of nominations around one or a few students (i.e., ``celebrities"~\cite{grunspan2016} who receive many more nominations than their peers). Finally, transitivity measures the tendency of nodes to cluster together and is calculated as the proportion of two-paths (two edges connecting three nodes) that have a third edge closing the triangle, not considering edge direction. If node A is connected to node B and node C, for example, an edge between nodes B and C would form a triangle. A higher proportion of such triangles would lead to higher transitivity values (i.e., closer to one).

We determined the standard errors of each of these statistics via bootstrapping: resampling the observed network many times, calculating the statistic of each sampled network, and then determining the standard deviation of the statistic among all of the sampled networks~\cite{traxler2020network,snijders1999non}. The bootstrapping was performed with 5,000 bootstrap trials for each network using the \textit{snowboot} package in R~\cite{chen2019snowboot}.

\subsubsection{Exponential random graph models}

We determined the extent to which a gender bias exists in each observed recognition network using exponential random graph models (ERGMs). Such models assume that an observed network is a realization from a random graph that comes from a distribution belonging to the exponential family~\cite{anderson1999,robins2007}. ERGMs allow us to perform many statistical tests at once, determining whether the frequencies of certain configurations (e.g., ties between students of the same gender) in our observed network are significantly different than if the ties were formed randomly. The goal is to use these $k$ configurations $g_k(y)$ and their corresponding coefficients $\theta_k$ to predict the formation of the random network $Y$. The model takes the form
\begin{equation*}
    P_\theta[Y = y] = \frac{\exp\left(\sum_{k} \theta_k g_k(y)\right)}{\sum_y \exp\left(\sum_{k}\theta_k g_k(y)\right)}
\end{equation*}
where $y$ is a realization of the random network $Y$ and the denominator serves as a normalization constant that ensures that the probability sums to one. Given an observed network $y$, the coefficients of the model are estimated using Maximum Likelihood Estimation (MLE). Due to the dependence between the network ties, the MLE is commonly approximated with Markov Chain Monte Carlo (MCMC) techniques~\cite{hunter2008}. The coefficients $\theta_k$ represent log-odds of tie formation and can be interpreted as a weighting of the importance of each modeled configuration for the realized network, where positive (negative) coefficients show that the configuration is observed more (less) frequently than by chance after accounting for all other configurations that are modeled. 

In our study, we fit an ERGM to each observed network using a similar set of configurations, or predictor variables, as our prior work~\cite{sundstrom2022perceptions}. For the lab course, we added two new variables. The first variable measured the tendency for students to nominate peers in their immediate lab group given prior work that suggests students often report connections to their group members on network surveys~\cite{pearson2017developing}. The second variable measured the tendency for students to nominate peers enrolled in their same lecture course (the separate course structure is different than in Ref.~\cite{sundstrom2022perceptions}). We also only measured discussion section homophily in the lecture course because our prior research found no significant tendency for students to nominate peers in their discussion section as strong in the lab material~\cite{sundstrom2022perceptions} and the discussion sections are now even further removed from lab material given the distinct courses. Different from our previous work, students received separate final course grades in the lab and lecture courses rather than one overall course grade that encompassed lab and lecture content. Therefore, we used the lab course final grades in the ERGMs for the lab course recognition networks and the lecture course final grades in the ERGMs for the lecture course recognition networks. Finally, we did not include a variable measuring transitivity in the models as we did in previous work~\cite{sundstrom2022perceptions} because the MCMC MLE did not converge with this variable added. The goodness-of-fit diagnostics, however, showed that our model sufficiently captured the distributions of indegree, outdegree (number of nominations reported by each student), and transitivity for all four observed networks (see Fig.~\ref{fig:gof} in the Appendix). The following predictor variables were included in our model:
\begin{itemize}
    \itemsep0em
    \item \textit{Edges}: intercept term equal to the number of edges in the network
    \item \textit{Reciprocity}: number of mutual nominations (i.e., student A nominates student B and student B nominates student A)
    \item \textit{Woman $\rightarrow$ woman}: number of edges for which a woman nominates another woman (base term is \textit{man $\rightarrow$ man})
    \item \textit{Woman $\rightarrow$ man}: number of edges for which a woman nominates a man (base term is \textit{man $\rightarrow$ man})
    \item \textit{Man $\rightarrow$ woman}: number of edges for which a man nominates a woman (base term is \textit{man $\rightarrow$ man})
    \item \textit{URM $\rightarrow$ URM}: number of edges for which a URM student nominates a URM student (base term is \textit{non-URM $\rightarrow$ non-URM})
    \item \textit{URM $\rightarrow$ non-URM}: number of edges for which a URM student nominates a non-URM student (base term is \textit{non-URM $\rightarrow$ non-URM})
    \item \textit{Non-URM $\rightarrow$ URM}: number of edges for which a non-URM student nominates a URM student (base term is \textit{non-URM $\rightarrow$ non-URM})
    \item \textit{Physics majors $\rightarrow$ physics majors} (lab course only): number of edges for which a student in the ``physics majors" lecture course nominates another student in the `` physics majors" lecture course (base term is \textit{non-majors $\rightarrow$ non-majors})
    \item \textit{Physics majors $\rightarrow$ non-majors} (lab course only): number of edges for which a student in the ``physics majors" lecture course nominates a student in the ``non-majors'' lecture course (base term is \textit{non-majors $\rightarrow$ non-majors})
    \item \textit{Non-majors $\rightarrow$ physics majors} (lab course only): number of edges for which a student in the ``non-majors" lecture course nominates a student in the ``physics majors" lecture course (base term is \textit{non-majors $\rightarrow$ non-majors})
    \item \textit{Lab group homophily} : number of edges connecting students in the same lab group
    \item \textit{Lab section homophily}: number of edges connecting students enrolled in the same lab section
    \item \textit{Discussion section homophily} (lecture course only): number of edges connecting students enrolled in the same discussion section
    \item \textit{Grade of nominee}: correlation between final course grade and number of received nominations
\end{itemize}
We used the coefficient estimates of the \textit{woman $\rightarrow$ woman}, \textit{woman $\rightarrow$ man}, and \textit{man $\rightarrow$ woman} variables for the four observed recognition networks to determine whether a gender bias exists in student nominations of strong peers after adjusting for the other network configurations included in the model. While we only focus on these gender variables in this paper, we keep the other variables in the model to account for as many different aspects of students' identity and participation in the course as possible and because an exploratory analysis indicated that removing these other variables can change the results for the gender variables~\cite{walsh2021omittedbias}. In particular, a few of the predictor variables (e.g., \textit{lab group homophily}, \textit{lab section homophily}, and \textit{discussion section homophily}) explicitly control for patterns of student interactions, allowing us to identify whether a gender bias in students' recognition of strong peers exists even after accounting for any strong interaction trends (e.g., gender homophily)~\cite{sundstrom2022interactions}. 

Additionally, we note that the final four predictor variables listed above (related to lab group, lab section, discussion section, and grade) cannot handle unknown data. Therefore, only students with known data for these four variables were included in the ERGM analysis. While this predominantly restricted analysis to students who completed the course (i.e., students who did not have a final course grade likely dropped or withdrew from the course after we administered the survey), the ERGM analysis still included more than 90\% of students that are part of our overall analysis. Thus, the statistical models provide an accurate description of most students in the class. We recommend for future work to investigate the network positionality of students who do not complete their physics course.

\begin{figure*}[t]
    \centering
    \includegraphics[width=6.6in]{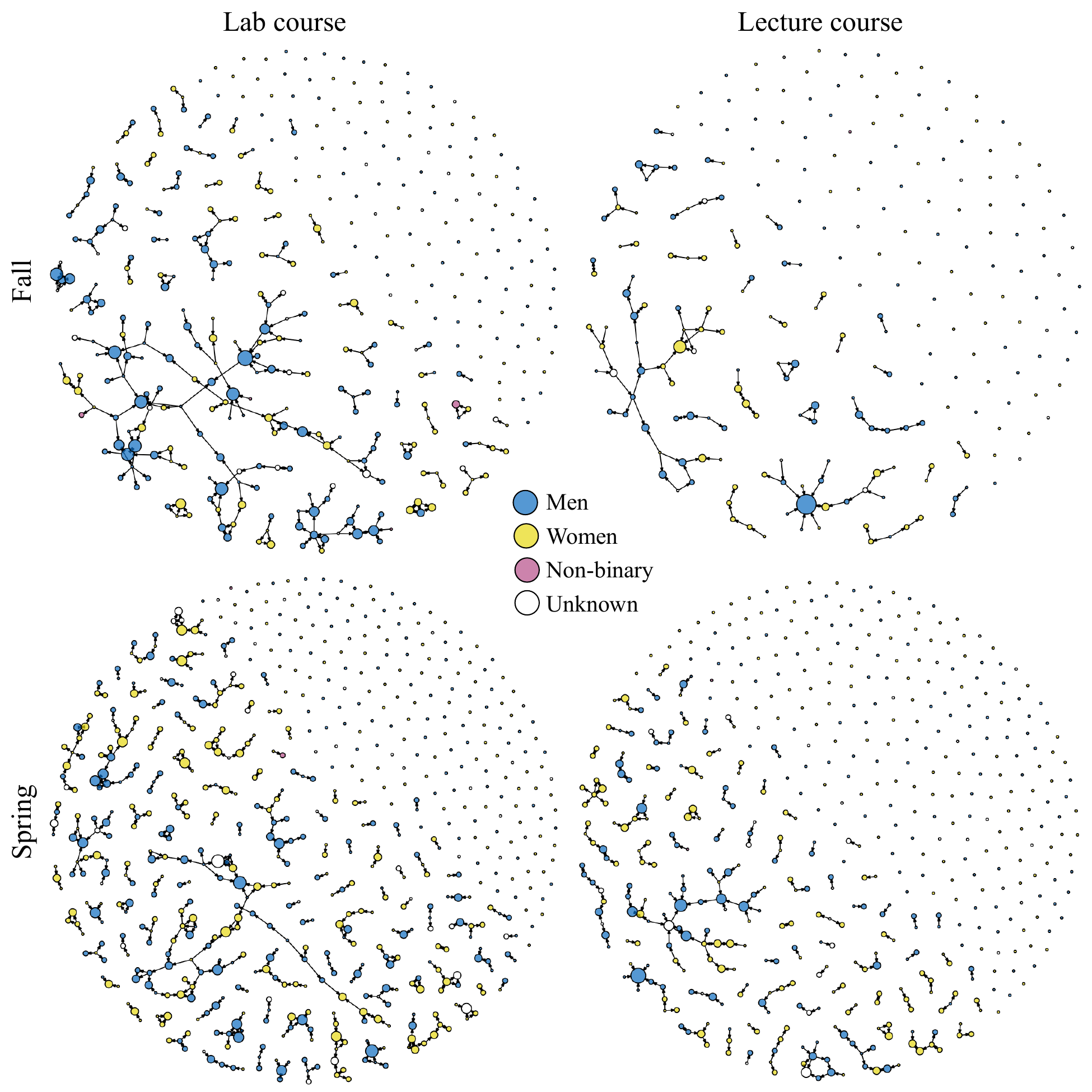}
    \caption{Recognition networks for all analyzed courses. Nodes are colored by gender and sized proportional to indegree (number of received nominations as strong in the course). Edges point from the nominator to the nominee.}
    \label{fig:gendersociograms}
\end{figure*}

We also note that some sample sizes, particularly for URM students, seem too small to make statistical comparisons with our models (Table~\ref{tab:demographics}). ERGMs, however, consider edges rather than nodes as the unit of analysis. Though the number of URM students (i.e., nodes) may be small, the networks we study include many of the possible edges between students (Table~\ref{tab:networkstats} and Fig.~\ref{fig:gendersociograms}). Smaller sample sizes, furthermore, do not prevent valid estimation of the coefficient values. Instead, they are reflected in the standard errors and \textit{p}-values of the coefficients~\cite{kolaczyk2015question}. Quantitative modifications to ERGMs are only necessary for very small networks (less than six nodes)~\cite{yin2021highly}.

We finally note that students' final grades in the lab course were fairly skewed, with many students earning an A or A-. This may introduce issues of range restriction for the \textit{grade of nominee} term, where low variability in students' final course grades limits the possibility of finding a significant correlation between grades and received nominations. We find in both offerings of the lab course, however, that the model is able to distinguish a significant effect of course grade on peer recognition (Table~\ref{tab:coefvalues}).

\begin{figure}[t]
    \centering
    \includegraphics[width=3.1in,trim={1cm 7cm 15cm 0 }]{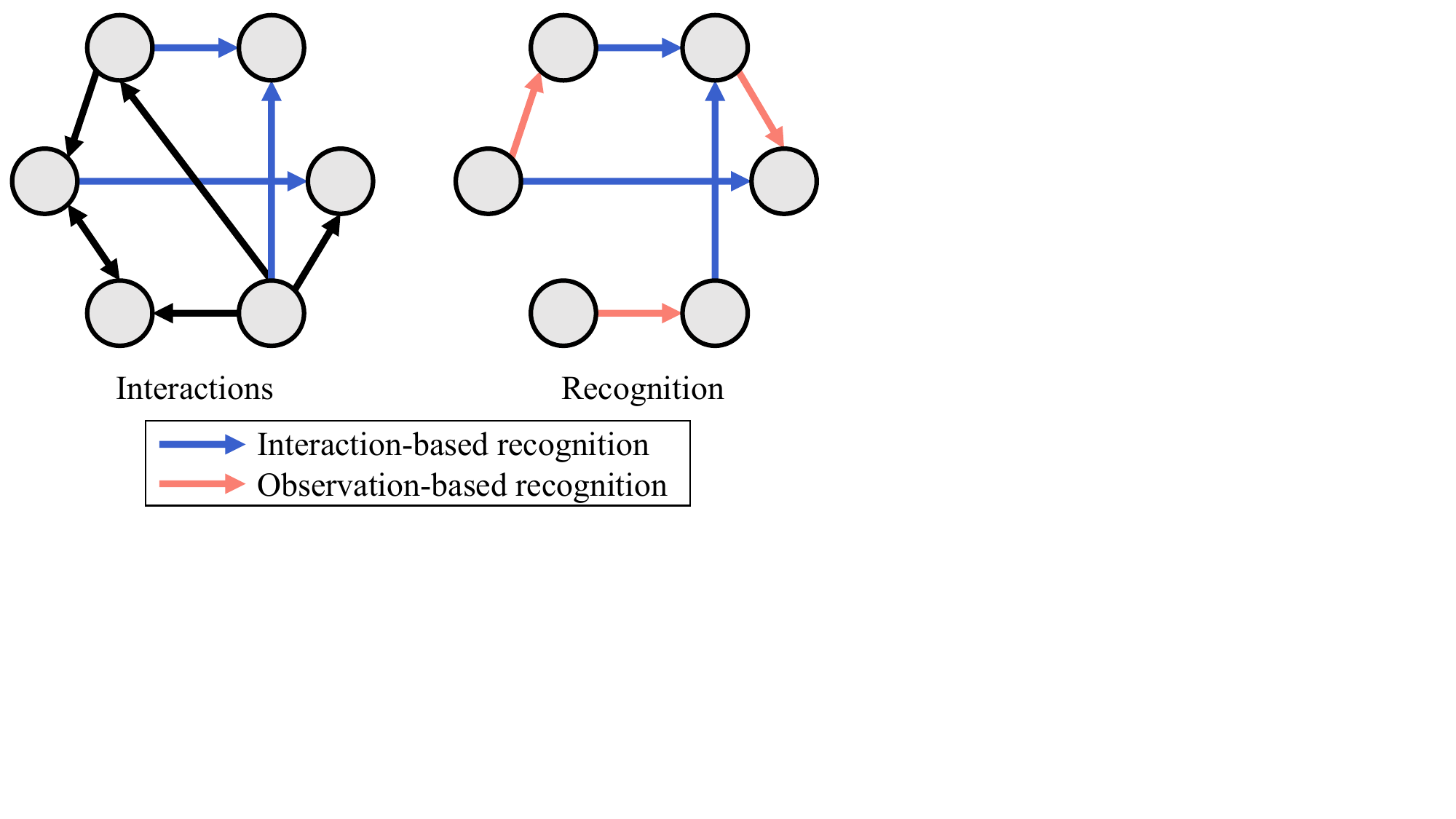}
    \caption{Toy interaction and recognition networks to exemplify how we calculated percent overlap and fraction of interaction network edges kept in the recognition network. Black edges indicate edges that appear in the interaction network but not the recognition network, blue edges indicate edges that appear in both the interaction and recognition networks (edges kept), and orange edges indicate edges that appear in the recognition network but not the interaction network. In this case, the percent overlap is $\frac{1}{2}$ because three out of six recognition network edges also appear in the interaction network. The fraction of interaction network edges kept in the recognition network is $\frac{1}{3}$ because three out of nine interaction network edges also appear in the recognition network.}
    \label{fig:schematic}
\end{figure}

\subsubsection{Peer interactions}
\label{sec:intanalysis}

To measure the extent to which peer interactions are related to peer recognition, we converted students' self-reported interactions (last two questions on the survey, see Fig.~\ref{fig:survey}) into directed networks. Similar to our analysis of the recognition network structure, we first calculated the density, the proportion of possible edges in the network that we observe, of each interaction network. For each offering of each course, we compared the recognition and interaction network densities to determine whether there were comparable numbers of edges in both networks or if one of the two networks contained many more edges than the other. This comparison was observational and not statistical.

For a more interpretable metric, we calculated the \textit{percent overlap}, the percent of directed edges in the recognition network that also appear in the corresponding interaction network (see Fig.~\ref{fig:schematic} for example), for each course. This measure allowed us to determine the extent to which peer recognition was \textit{interaction-based} or \textit{observation-based}. Higher percent overlap values indicate that most of students' nominations of strong peers were interaction-based recognition: many students nominated peers with whom they also reported interacting. For interaction-based recognition, we assume that the nominator came to understand the nominee's skill set through their direct interactions (e.g., talking to immediate group mates during lab). Lower percent overlap values, on the other hand, indicate that most of students' nominations of strong peers were observation-based recognition: many students nominated peers with whom they did not report interacting. For observation-based recognition, we assume that the nominator came to understand the nominee's skill set through their indirect observations of them (e.g., seeing someone frequently participate in lecture) rather than direct interactions.

We also calculated \textit{gender homophily} as the percent of edges in the interaction network where both the nominator and the nominee are of the same gender (i.e., edges from men to men and edges from women to women). Research has shown that gender homophily is prevalent in student interaction networks~\cite{dokuka2020academic,sundstrom2022interactions,mcpherson2001birds}, thus this measure helped us understand patterns between the interaction and recognition networks related to gender, discussed next.

Finally, we compared the extent to which any gender bias in peer recognition was specifically related to a gender bias in interaction-based recognition (i.e., if there was a gender bias in which of students' interaction ties they also nominated as strong in the course). This analysis only included edges for which both the nominator and the nominee self-reported their gender as either man or woman (859 out of 1,000 total nominations of strong peers and 1,590 out of 1,789 total self-reported interactions across all four courses). For every possible combination of men and women nominating each other (i.e., man nominating a man, man nominating a woman, woman nominating a man, and woman nominating a woman), we calculated the \textit{fraction of interaction network edges kept in the recognition network} as the number of directed edges that appear in both the interaction and recognition networks divided by the number of directed edges in the interaction network (see Fig.~\ref{fig:schematic} for example). We compared this measure by student gender in each network. We did not perform statistical tests because the goal of this analysis was to determine large-scale trends in the measure and relying on $p$-values can be problematic~\cite{cohen1994earth, cumming2013understanding, nosek2018preregistration}. Statistical tests of distinguishability would involve many comparisons that increase the risk of finding apparent statistical significance due only to chance. Instead, we use overlap in error bars (given by standard errors) to make qualitative interpretations about differences in the measure between men and women and do not comment on the possible distinguishability of small effects. While this approach is more appropriate than statistical testing, we acknowledge that using error bars may come with its own set of limitations~\cite{correll2014error}.

%This measure captured the extent to which students recognized peers with whom they interacted as strong in the course material and provides information about any biases in which peers students choose to include from their interaction network in their recognition network.

%Observation-based recognition measured the extent to which students recognized peers with whom they did not interact, but likely indirectly observed, as strong in the course material and was calculated as the number of edges in the recognition network that did not appear in the interaction network divided by the total number of edges in the interaction network. This measure provides information about any biases in which peers students choose to recognize based on indirect observations, essentially measuring the percent increase in the number of edges in the interaction network when forming the recognition network.

We note that we could not measure the extent to which any gender bias in peer recognition was related to a gender bias in observation-based recognition. An appropriate measure of such a bias would be to calculate the fraction of peers that students indirectly observed, but did not interact with, that they nominated as strong in the course material. However, we did not collect data about which peers students indirectly observed -- all we know is who they do not interact with. In large classes, such as the ones we analyze here, it is not reasonable to assume students indirectly observed all of the peers with whom they did not interact. We recommend for future work to investigate gender bias in observation-based recognition by examining peer recognition in small courses (where it is reasonable to assume students have the chance to indirectly observe all of their peers) or by also collecting data about peers with whom students are familiar but did not directly interact.

%We will also show that interacting with peers did not fully explain the source of peer recognition: a little less than half of the nominations of strong peers did not coincide with an edge in the interaction network (percent overlap of roughly 50\%) and XX. Therefore, we performed a qualitative analysis of students' written explanations for nominating peers as strong in the course.

\subsubsection{Explanations}

We conducted a thematic coding analysis of student responses to the survey prompt, ``Please briefly explain why you chose this student as strong in the course material," to identify what gets recognized in peer recognition -- that is, the skill sets for which students recognized their strong physics peers. The first author initially read all responses to gain a sense of the data as a whole~\cite{Tesch1990}. Upon recognizing similar themes to those identified in prior research~\cite{tonso2006student,danielsson2012exploring,fields2013picking,due2014competent,gonsalves2014physics,gonsalves2014persistent,gonsalves2016masculinities,irvingsayre2015,DoucetteHermione,doucettegoodlabpartner,cooper2018perceives,stump2022perc}, the first author drafted an \textit{a priori} codebook informed by these themes. The research team then iteratively developed this codebook by individually coding a subset of the data and then meeting to modify the code definitions based on coding disagreements~\cite{Campbell2013}. We also grouped together similar codes into four overarching categories: knowledge, processes, interactions, and other. These categories were inspired by those in Ref.~\cite{adams2015analyzing}, in which the authors categorized students' problem solving skills as related to knowledge, processes, and beliefs.

After the coding scheme was agreed upon, three members of the research team coded a stratified random sample of 10\% of the 1,000 total explanations in our data set. We stratified the random sample by course (lab and lecture) because the two courses had different learning objectives and course structures, which might have led students to associate different skill sets with being strong in each course. Therefore, half of the random sample contained explanations from the lab course and the other half contained explanations from the lecture course. We determined interrater reliability by calculating Fuzzy Kappa~\cite{kirilenko2016inter} between each of the three pairs of coders because each explanation could receive multiple codes. All three pairwise Fuzzy Kappa values were greater than 0.8 indicating sufficient interrater reliability~\cite{kirilenko2016inter}. After reaching this level of agreement, the first author coded the remaining explanations.

We then compared the fractions of nominations between every possible combination of men and women nominating each other (i.e., man nominating a man, man nominating a woman, woman nominating a man, and woman nominating a woman) containing each code. This comparison only included explanations for which both the nominator and the nominee self-reported their gender as either man or woman (859 out of the 1,000 total explanations across all four courses). We also aggregated the data from the fall and spring offerings because the results for each individual offering were not substantially different from the aggregated results and the larger data set reduces possible statistical noise. Similar to our explanations analysis, we did not perform any statistical tests because the goal of this analysis was to identify any large-scale differences in these fractions. Instead, we used overlap in error bars (given by standard errors) to make qualitative interpretations about differences in code frequencies between men and women. This comparison allowed us to determine whether and how any gender bias in peer recognition in each course is related to what gets recognized (i.e., students nominating one another for different skill sets).

\section{Results}

We present the results from each of the four stages of analysis (Fig.~\ref{fig:methods}): recognition network structure, ERGM analysis, comparison to interaction networks, and explanations of nominations of strong peers.

\begin{table}[b] 
\centering
\caption{\label{tab:networkstats}
Network-level statistics for the observed recognition networks. Standard errors of the last digit are shown in parentheses.}
\begin{ruledtabular}
\setlength{\extrarowheight}{1pt}
\begin{tabular}{lcccc}
 & \multicolumn{2}{c}{Lab course} &  \multicolumn{2}{c}{Lecture course}  \\ 
 \cline{2-3}
 \cline{4-5}
 & Fall & Spring & Fall & Spring \\ 
 \hline
 Nodes & 387 & 646 & 237 & 513  \\ 
Density & 0.002(2) & 0.0009(8) & 0.002(3) & 0.0008(9)  \\ 
Indegree centralization & 0.011(4) & 0.005(2) & 0.03(1) & 0.009(3)  \\
Transitivity & 0.24(3) & 0.26(2) & 0.13(3) & 0.11(2)  \\
%Proportion in giant component & 0.23(9) & 0.07(8) & 0.13(9)  & 0.07(6)  \\
%Proportion isolates & 0.33(4) & 0.32(3) & 0.47(5) & 0.48(3)  \\
\end{tabular} 
\end{ruledtabular}
\end{table}

\subsection{Recognition network structure}

The structural features of the observed recognition networks, summarized in Table~\ref{tab:networkstats} and shown in Fig.~\ref{fig:gendersociograms}, provide information about broad patterns of recognition. %We cannot directly compare the density or indegree centralization values across the observed networks because these measures do not scale linearly with the number of nodes in the network. 
We observe that the densities of the lab and lecture recognition networks are similar within each semester (fall and spring). Because there are many more nodes in the lab recognition network than the lecture recognition network in each semester (Table~\ref{tab:networkstats}), the densities suggest that there is a higher level of connectedness in the lab networks than the lecture networks (i.e., a higher proportion of possible edges exist in the networks in the left column of Fig.~\ref{fig:gendersociograms} than in the corresponding network in the right column). We also observe relatively low indegree centralization values in each network. This observation indicates that the nominations are fairly spread out among the students in a course rather than concentrated around only one or a few students. Correspondingly, there are no outstanding ``celebrities" in any network; no nodes are much larger than the rest, which would be associated with receiving many nominations (Fig.~\ref{fig:gendersociograms}). While there is one man in the fall lecture network who receives more nominations than anyone else (seven), outstanding celebrities in the large science courses analyzed in prior work receive more than 30 nominations~\cite{grunspan2016,sundstrom2022perceptions}.

We also observe that all four networks contain one or two relatively large components that connect many nodes along chain-like formations and many smaller components of two to four nodes that are only connected to each other. The prevalence of these smaller components, however, is stronger in the lab course than the lecture course as indicated by the higher levels of transitivity. This pattern is likely due to the lab course placing more emphasis on small group work during the lab sessions (e.g., coordinating experimental investigations and submitting the lab notes for a group grade) than the lecture course does during the discussion sessions (e.g., collaborating on problems but not submitting work for a grade). We also see a large fraction of isolated nodes (at least 30\% of the total nodes in each network), representing individuals who responded to the survey but did not nominate any peers as being strong in the course material and who were also not nominated by any other students. The demographics of the isolated nodes (e.g., gender and race or ethnicity) in each network are proportional to the demographics of the course population.

\begin{figure}[t]
    \centering
    \includegraphics[width=3.2in,trim={1cm 1cm 12.3cm 0}]{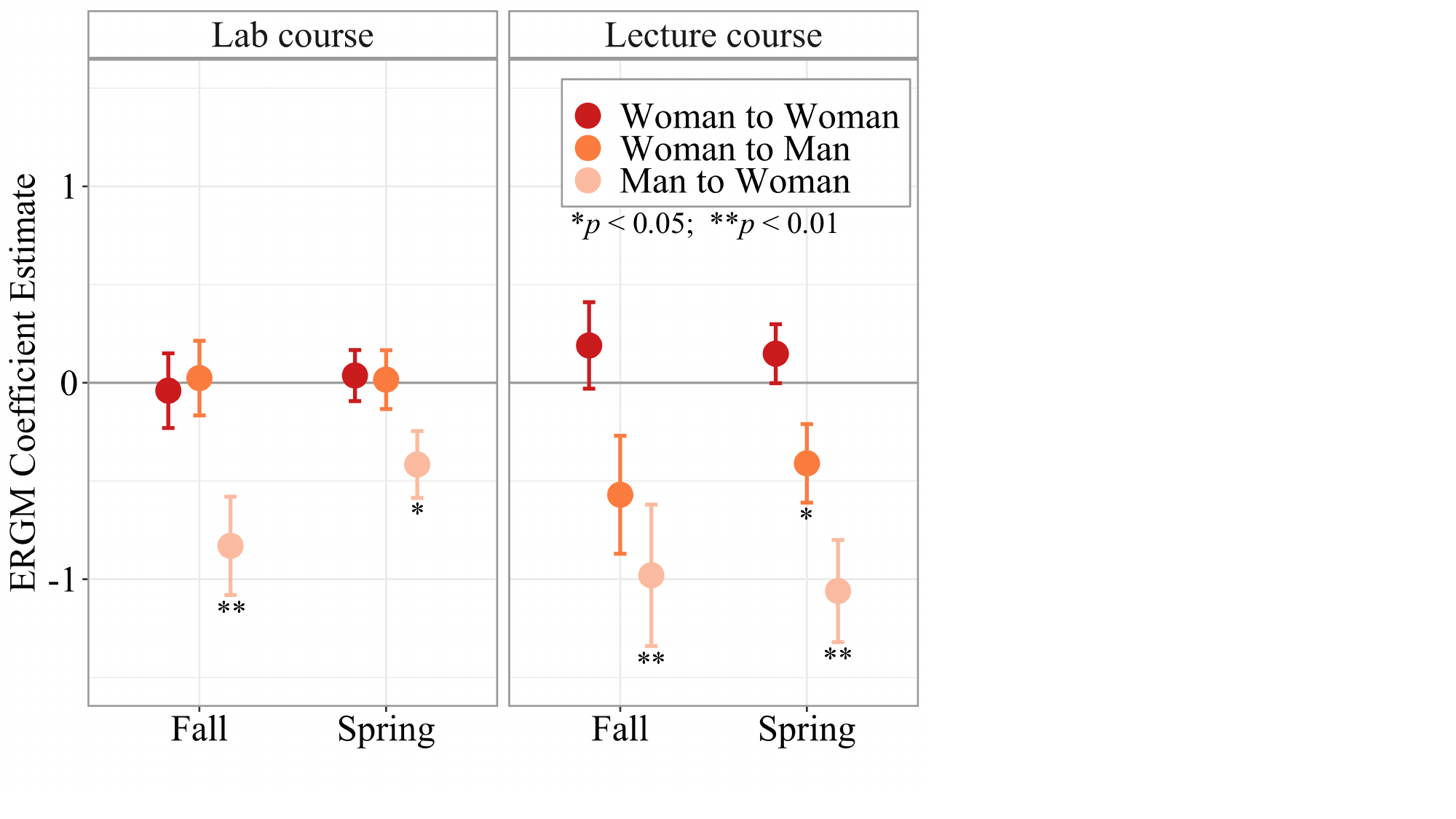}
    \caption{Coefficient estimates, represented as log-odds, for the gender variables of the exponential random graph models (values shown in Table \ref{tab:coefvalues} in the Appendix). The base term (i.e., coefficient estimate of zero) is nominations from man to man. Error bars represent standard errors and asterisks indicate statistical significance.}
    \label{fig:genderergm}
\end{figure}

\subsection{Exponential random graph models}

As per our research questions, we focus on the coefficient estimates of the ERGM terms that speak to the extent to which a gender bias exists in the recognition networks (see Fig.~\ref{fig:genderergm} and Table~\ref{tab:coefvalues} in the Appendix). We find that in both offerings of the lab course, women proportionately nominate men and women as strong in the course material as compared to men nominating men (red and orange dots on the left panel of Fig.~\ref{fig:genderergm}). Women also proportionately nominate men and women as strong in the course material as compared to men nominating men in the fall offering of the lecture course (orange dots on the right panel of Fig.~\ref{fig:genderergm}). Women nominate men less frequently than men nominate men, however, in the spring offering of the lecture course. 
%We note that this tendency is only statistically significant in the spring offering of the lecture course, but the coefficient estimate for the fall offering is comparable in magnitude and likely indicates a similar pattern.

Additionally, in both offerings of both courses men disproportionately under-nominate women as strong in the course material as compared to men nominating men (pink dots on both panels of Fig.~\ref{fig:genderergm}). Although comparing ERGM coefficient values across different-sized networks is ill-defined~\cite{duxbury2021problem}, we tentatively observe that this bias from men occurs to a similar extent in every course, with the possible exception of the spring offering of the lab course which has a slightly smaller coefficient estimate for the \textit{man $\rightarrow$ woman} variable. %These statistical results related to gender are also visible in the network diagrams (Fig.~\ref{fig:gendersociograms}), where we see that men on average receive more nominations as strong in the course than women in every network (the average size of blue nodes is larger than the average size of yellow nodes in each network).

In all cases, the comparisons are made after adjusting for the other variables in our model. We particularly note that these results related to gender bias hold even after controlling for the gender composition of lab groups (\textit{lab group homophily} variable), which were intentionally made to avoid isolated women, and other patterns of student interactions (e.g., \textit{lab section homophily} and \textit{discussion section homophily} variables). In the lab course, the results also hold after controlling for any bias based on lecture course enrollment (\textit{physics majors $\rightarrow$ physics majors}, \textit{physics majors $\rightarrow$ non-majors}, and \textit{non-majors $\rightarrow$ physics majors} variables). While the non-majors lecture course is fairly gender balanced (Table~\ref{tab:demographics}), the physics majors course contains a majority of men (70-80\%, depending on the semester). A bias in which nominations favor students in the physics majors course, therefore, would make men more likely to be nominated than women. The gender bias we observe, however, is present even after controlling for the lecture enrollment variables included in our model (and thus the different student populations in the two lecture courses).

\begin{table}[t] 
\centering
\caption{\label{tab:interactionstats}
Densities and gender homophily of the observed interaction networks and percent overlap of the recognition and interaction networks. Standard errors of the last digit of the densities are shown in parentheses.% and asterisks indicate statistical significance ($^{***}p<0.001$).
}
\begin{ruledtabular}
\setlength{\extrarowheight}{1pt}
\begin{tabular}{lcccc}
 & \multicolumn{2}{c}{Lab course} &  \multicolumn{2}{c}{Lecture course}  \\ 
 \cline{2-3}
 \cline{4-5}
 & Fall & Spring  & Fall  & Spring  \\ 
 \hline
 Density & 0.003(3) & 0.002(1) & 0.005(5) & 0.0015(9)  \\ 
%  Correlation with perception network & 0.40$^{***}$ & 0.43$^{***}$ & 0.37$^{***}$ & 0.41$^{***}$  \\ 
 %QAP coefficient estimate & 6.29$^{***}$ & 7.07$^{***}$ & 5.94$^{***}$ & 7.06$^{***}$ \\
 %Odds ratio & 540 & 1,177 & 38 & 1,166\\
 Percent overlap & 56\% & 54\% & 57\% & 55\%  \\
 Gender homophily & 67\% & 67\% & 69\% & 72\% 
\end{tabular} 
\end{ruledtabular}
\end{table}

\subsection{Peer interactions}

To further understand how students identify who to recognize as strong in their physics course, we also analyzed students' self-reported interactions. The interaction network diagrams are shown in the Appendix (see Fig.~\ref{fig:genderintsociograms}).

Comparing the densities of the interaction networks (Table~\ref{tab:interactionstats}) to the densities of the corresponding recognition networks (Table~\ref{tab:networkstats}), we find in all four courses that the interaction network is more dense (i.e., students are more connected by edges) than the recognition network. This comparison may suggest that when making their nominations of strong peers, students select a subset of the peers with whom they interact to nominate as strong in the course material.

The percent overlap values, the percent of edges in the recognition network that are also in the corresponding interaction network, however, indicate that interactions only account for a little more than half of the nominations of strong peers in each recognition network (Table~\ref{tab:interactionstats}). That is, students also nominate peers with whom they did not report interacting. These results indicate that interaction-based recognition (e.g., learning about peers' skills by working together on a problem set or discussing concepts together in lecture) and observation-based recognition (e.g., learning about peers' skills by seeing a student ask a question in class or watching a student in a nearby lab group collect data for an experiment) occur with similar frequencies in the observed courses.

\begin{figure}[t]
    \centering
    \includegraphics[width=2.4in,trim={1cm 0cm 22cm 0}]{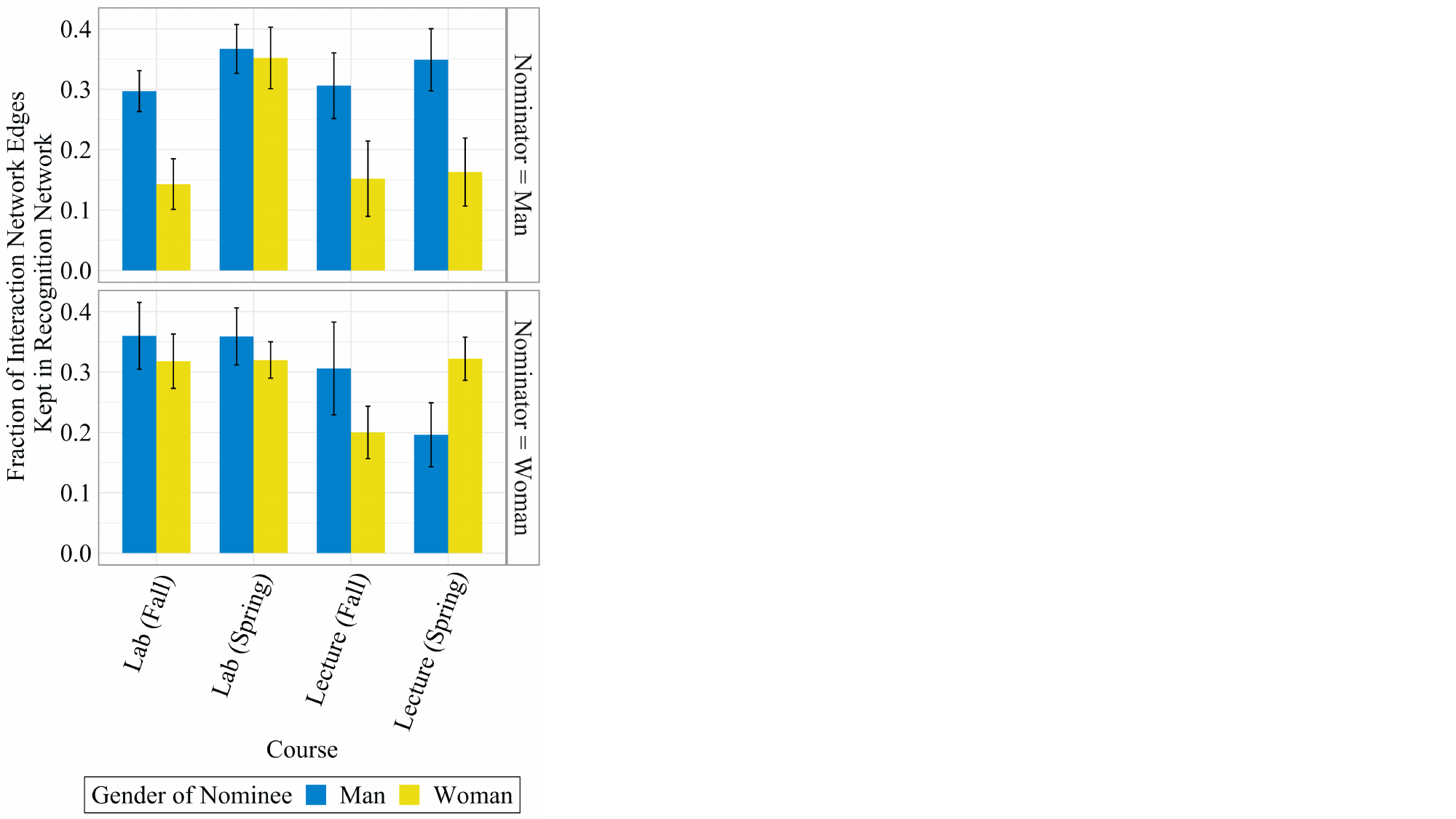}
    \caption{Fraction of edges in each interaction network that also appear in the corresponding recognition network for each combination of nominator and nominee gender. Error bars represent standard errors of the proportions.
    }
    \label{fig:addrop}
\end{figure}

We examined the extent to which the gender biases in peer recognition identified in the ERGMs (in which men under-nominate women in all courses and women under-nominate men in the spring lecture course) are related to gender biases in the interaction-based nominations. We identified the fraction of students' interaction ties that they also nominate as strong in the course, split by gender of the nominee (Fig.~\ref{fig:addrop}). In three out of the four courses (all but the spring offering of the lab course), we observe that men disproportionately ``keep" more of their interaction ties with men than with women (see top panel of Fig.~\ref{fig:addrop}). This pattern indicates that the gender bias coming from men that we observed in the ERGM analysis for these courses is (at least partly) related to a gender bias in interaction-based recognition. Men exhibit strong strong gender homophily in their peer interactions (Table~\ref{tab:interactionstats}) and, after adjusting for this homophily (i.e., the metric shown in Fig.~\ref{fig:addrop} is normalized by the number of interactions made to peers of a given gender), men exhibit a gender bias in which of their interaction ties they select as strong in the course material. 

%That is, the analysis assumes the initial gender homophily in peer interactions and demonstrates further gender bias in how men select students from their interaction networks to nominate as strong in the course. %  because the fractions in Fig.~\ref{fig:addrop} are calculated within each combination of nominator and nominee gender separately (e.g., we compare how many of men's interaction edges to men they also nominate as strong in the course and, separately, how many of men's interaction edges to women they also nominate as strong in the course).

In the spring offering of the lab course, in contrast, men proportionately nominate their interaction ties to men and women as strong in the course (see top panel of Fig.~\ref{fig:addrop}). The gender bias in peer recognition coming from men in this course, therefore, is likely related to a gender bias in observation-based recognition, though we could not measure such a bias in our study (see Sec.~\ref{sec:intanalysis}).

Finally, the gender bias in the spring lecture course, in which women under-nominate men as strong in the course, is (at least partly) related to interaction-based recognition: women disproportionately nominate more of their interaction ties with women than with men (see bottom panel of Fig.~\ref{fig:addrop}). In all other semesters, where we do not observe a gender bias coming from women in the ERGMs, we correspondingly find that women nominate proportional numbers of men and women from their interaction ties as strong in the course material.

\begin{table*}[t]
\caption{\label{tab:codingscheme}%
Definitions and examples of our coding scheme for students' explanations of why they nominated their peers as strong in the course material. Some codes were only present in either lab or lecture course nominations, indicated in parentheses. \textit{N} indicates the total number of occurrences of each code in each course, lab or lecture.}
\begin{ruledtabular}
\setlength{\extrarowheight}{1.2pt}
\begin{tabular}{p{0.24\linewidth}  p{0.28\linewidth} p{0.32\linewidth}p{0.04\linewidth} p{0.04\linewidth}}
Category or code & Definition & Example & \textit{N}$_{\text{lab}}$ & \textit{N}$_{\text{lecture}}$\\
\hline
Knowledge \\
\hspace{3mm}Understanding & Knowledgeable about the course material & \hangindent=1em``Seems to have a strong sense of the topics in class." & 228 & 112 \\
\hspace{3mm}Performance& \hangindent=1em Receiving good grades; answering questions correctly & “Did well on the quiz.” & 16 & 38\\
\hspace{3mm}Experience & \hangindent=1em Having relevant background knowledge or experiences outside of the course  & \hangindent=1em “Previous understanding of the class from AP Physics.” & 36 & 16\\
\hspace{3mm}Natural ability & \hangindent=1em Having an innate aptitude for understanding the course material &  \hangindent=1em “Has an innate ability to view a problem in its simplest terms.” & 6 & 12\\
\hspace{3mm}Motivation & \hangindent=1em Putting a lot of time or effort into the course; determined &  \hangindent=1em “Works harder than anyone I know to improve their physics lab knowledge.” & 88 & 44\\
Processes \\
\hspace{3mm}Analysis (lab only) & \hangindent=1em Analyzing and interpreting experimental data &  \hangindent=1em “Has great understanding of software to help analyze data.” & 148 & n/a\\
\hspace{3mm}Planning (lab only) & \hangindent=1em Designing and evaluating an experimental procedure &  \hangindent=1em “Knows a lot about experimental design and what to do to answer the experimental question.”& 90 & n/a\\
\hspace{3mm}Data collection (lab only) & \hangindent=1em Carrying out an experimental procedure  & \hangindent=1em ``Good at experimental setup/conducting trials with relatively small errors when possible." & 26 & n/a \\
\hspace{3mm}Writing (lab only) & \hangindent=1em Writing lab notes about an experiment &  “Good at making detailed write-ups.” & 21 & n/a\\
\hspace{3mm}Problem solving (lecture only) & \hangindent=1em Visualizing or reasoning through problems; applying the right equations to problems  & \hangindent=1em ``Very good at identifying the topics that a problem contains and quickly connecting it with a formula." & n/a & 54\\
Interactions \\
\hspace{3mm}Helping & \hangindent=1em Providing support with the course material to others (nominator mentions benefits from this support) &  “Helped me understand the homework.” & 77 & 58
\\
\hspace{3mm}Explaining & \hangindent=1em Describing or clarifying the course material to others (nominator does not mention benefits from this explaining) &  ``Explained concepts to me very well." & 23 & 31 \\
\hspace{3mm}Participation & \hangindent=1em Active contributor to in-class discussions and activities; asking questions & “Participates effectively during the lecture.” & 54 & 7\\
\hspace{3mm}Leading & Taking charge during in-class group work &  \hangindent=1em “Organizes the group quite well and gets the group rolling, keeping us on track throughout the lab.” & 17 & 2\\
Other \\
\hspace{3mm}Other &  \hangindent=1em Part or all of the explanation is vague or does not fit with above codes$^{\dag}$ & ``Thinks outside the box." & 52 & 22\\
\hspace{3mm}None & No explanation provided & [Blank] & 57 & 26\\
\end{tabular}
\end{ruledtabular}
\raggedright $^{\dag}$ A more detailed description of explanations coded as Other, with examples, is provided in the Appendix.
\end{table*}

\subsection{Explanations}

We devised a coding scheme characterizing students' written explanations of their nominations of strong peers to determine what gets recognized in peer recognition -- that is, what are the specific skill sets that students associate with being strong in their physics course (see Table \ref{tab:codingscheme}). The coding scheme illuminates the features of peer interactions and indirect observations of peers that students consider when selecting which of their peers to nominate. 

Related to knowledge, students describe strong peers as those who have \textit{understanding} of the course material and have high \textit{performance} (e.g., earn high grades). Students also mention that the peers they nominate have \textit{experience} or background knowledge relevant to the course and have a \textit{natural ability} for learning physics. Nominees were also described as hard-working or having \textit{motivation} to learn the course material.

Students identify multiple processes associated with being strong in their physics courses. These codes are course-specific. In the lab course, students describe strong peers as those who carried out the data \textit{analysis} for their experiment, contributed to the \textit{planning} or experimental design, participated in \textit{data collection}, and engaged in \textit{writing} up the lab notes. In the lecture course, students acknowledge the \textit{problem solving} abilities of their strong peers.

Students also consider features of their interactions with others in the course when forming perceptions of their peers. Some students describe strong peers as \textit{helping} them learn the course material and \textit{explaining} the course material to others. Students also describe nominees as having high levels of verbal \textit{participation} during lectures or group work and \textit{leading} group work during in-class activities.

\begin{figure*}[t]
    \centering
    \includegraphics[width=6.5in,trim={1.5cm 3.5cm 3cm 0}]{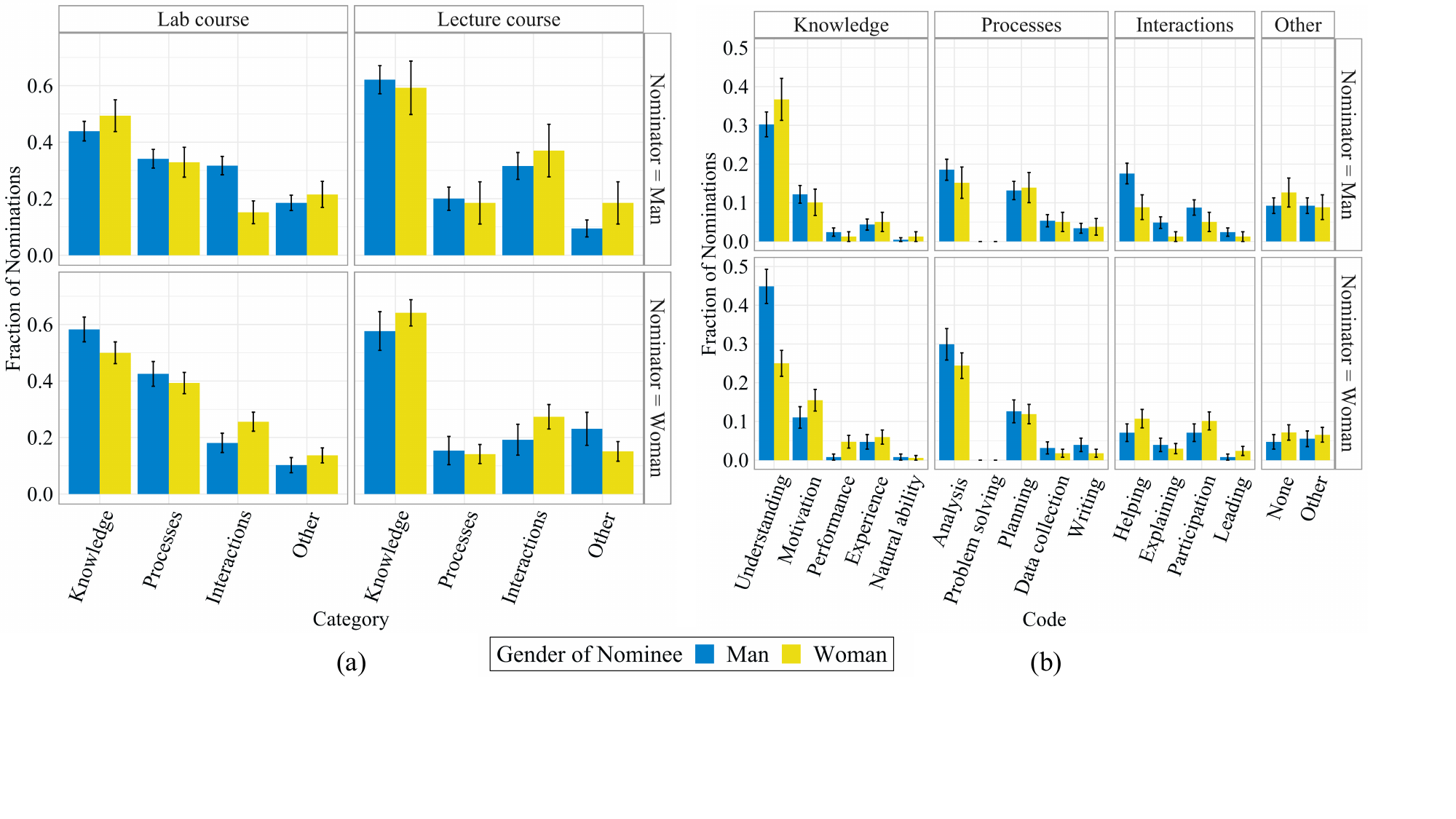}
    \caption{Fraction of nominations within each gender combination falling under each (a) category of our coding scheme within each course and (b) code of our coding scheme for the \textit{lab course only}, split by gender of the nominator and nominee. Results are aggregated across the fall and spring offerings of each course.  We coded 300 (205 in lab and 95 in lecture), 106 (79 in lab and 27 in lecture), 179 (127 in lab and 52 in lecture), and 274 (168 in lab and 106 in lecture) explanations from man to man, man to woman, woman to man, and woman to woman, respectively. Fractions do not necessarily sum to one because each explanation could receive multiple codes. Nominations made by men and women received comparable numbers of codes: men's nominations received an average of 1.4 codes and women's nominations received an average of 1.3 codes. Error bars represent standard errors of the proportions. 
    }
    \label{fig:expbygender}
\end{figure*}

Finally, some students note \textit{other} reasons that they viewed their peers as strong in the course material. These explanations are either too vague to associate with one of the above codes or too infrequent to create a separate code. Explanations left completely blank are coded as \textit{none}. 

Students proportionately recognize their interaction-based and observation-based nominees for each of these skills, except for process-related codes. In the lab course, students are more likely to describe their observation-based nominees than their interaction-based nominees as being strong in processes (specifically, data analysis). In the lecture course, students are more likely to describe their interaction-based nominees than their observation-based nominees as being strong in processes (specifically, problem solving). More detail about this comparison can be found in the Appendix (see Fig.~\ref{fig:explanationsappendix}).

Comparing fractions of nominations containing each code by the nominator and nominee gender, we find that in both the lab and lecture courses women proportionately nominate men and women for each category of the coding scheme (bottom row of Fig.~\ref{fig:expbygender}a), with small differences for most individual codes (bottom row of Fig.~\ref{fig:expbygender}b). One interesting exception is that women disproportionately nominate men more than women for \textit{understanding} in the lab course, though women do not exhibit a gender bias in their nominations in either offering of this course (Fig.~\ref{fig:genderergm}).

In the lecture course, men also proportionately nominate men and women for each category despite the gender bias identified in the ERGMs (top right box of Fig.~\ref{fig:expbygender}a). In the lab course, however, men nominate men more than women for features of their interactions (top left box of Fig.~\ref{fig:expbygender}a). Examining the individual codes, we see that men disproportionately over-nominate men as strong in the lab course for two of the individual interactions codes (top row of Fig.~\ref{fig:expbygender}b) -- \textit{helping} and \textit{explaining}, -- with the largest difference occurring for \textit{helping}. It is important to note that these comparisons for individual codes are limited by small sample sizes (Table~\ref{tab:codingscheme}). 

These results add nuance to our previous stages of analysis. The gender bias in the lecture course is seemingly related to who gets recognized (i.e., students disproportionately over-nominating interaction ties to peers of their same gender) and not what gets recognized. In the lab course, however, the observed gender bias is also due to what gets recognized: men nominate more men than women because of the ways they interacted.

\section{Discussion}

In this study, we aimed to disentangle who and what gets recognized in peer recognition to better understand the nature of gender bias in such recognition (identified in, e.g., Refs.~\cite{grunspan2016,bloodhart2020,sundstrom2022perceptions}). Across two offerings of distinct lab and lecture physics courses, we found that students determine who gets recognized in peer recognition in two ways, each with a similar frequency: interacting with peers and indirectly observing peers with whom they do not interact. We also identified what gets recognized in peer recognition: students mention skill sets related to knowledge, processes, and interactions in their written explanations of nominations. 

In the following sections, we synthesize these results related to gender bias in peer recognition (in which men disproportionately under-nominated women as strong in both courses and women disproportionately under-nominated men as strong in one offering of the lecture course) and relate our findings to prior work. We also discuss other implications for research suggested by our analysis.

\subsection{The nature of gender bias in peer recognition}

In both offerings of the lecture course, we found a similar gender bias in peer recognition (in which men under-nominated women compared to men) to previous work examining science courses aimed at first-year students during in-person and remote physics courses~\cite{sundstrom2022perceptions,bloodhart2020} and in-person biology courses~\cite{grunspan2016}. Such a bias was not previously observed in in-person mechanical engineering courses~\cite{salehi2019} and remote physics courses~\cite{sundstrom2022perceptions} aimed at beyond-first-year students, which is likely due to students in those courses being more familiar with each other, as noted in Ref.~\cite{sundstrom2022perceptions}. Different from prior work~\cite{grunspan2016,bloodhart2020,sundstrom2022perceptions}, we also observed a bias from women in the spring offering of the lecture course: women disproportionately under-nominated men as strong in this course. Surprisingly, we also found a gender bias in peer recognition (in which men disproportionately under-nominated women) in both offerings of the lab course even though we did not previously observe a gender bias within a comparable lab context of a remote physics course (Course A in Ref.~\cite{sundstrom2022perceptions}). We build on the body of previous work by also probing the nature of these gender biases in peer recognition, finding that whether the gender bias is related to who and/or what gets recognized varies by instructional context, whether lecture or lab, described next. 

\subsubsection{Gender bias in the lecture course is related to who gets recognized, not what gets recognized}

Our analysis suggests that in both offerings of the lecture course, the gender bias in peer recognition coming from men's nominations is related to men disproportionately over-nominating men with whom they interact as compared to women with whom they interact as strong in the course material (i.e., there was a gender bias in interaction-based recognition). We observed that this bias is likely \textit{not} attributable to men recognizing men and women for different skill sets because, of the people they recognize, men nominated men and women for similar skill sets. 

The latter result is somewhat surprising given the research literature showing that students associate men and women with having different skill sets in lecture~\cite{gonsalves2014persistent,due2014competent,doucettegoodlabpartner}. One study, for example, found that students associate men more than women with having a natural ability for learning physics and associate women more than men with asking questions~\cite{due2014competent}. If our observed gender bias in the lecture course is instead due to who gets recognized, rather than what gets recognized, then the observed bias may be due to gender stereotypes more broadly (i.e., students generally associating men more than women with being strong in physics)~\cite{hasse2002gender,danielsson2012exploring,gonsalves2016masculinities,kessels2006goes,makarova2015trapped,makarova2019,tate2005does,ceglie2011underrepresentation,moss2012,eddy2014,carlone2007understanding,kalender2019gendered,avraamidou2022identities}. Alternatively, this gender bias may be related to prior literature's suggestion that students' social networks, and subsequently peer perceptions, exhibit strong gender homophily~\cite{dokuka2020academic,sundstrom2022interactions,mcpherson2001birds}. This interpretation is supported by our analysis of women's nominations, described next.

We found that women under-nominated men in the spring offering of the lecture course, \textit{opposite} of what prior work would predict (i.e., a gender bias in peer recognition in favor of men, rather than against men)~\cite{grunspan2016,bloodhart2020,sundstrom2022perceptions}. In this offering, the observed bias was related to women nominating more of their interaction edges to women than men as strong in the course. Similar to nominations from men in this course, women proportionately nominated men and women for the skills identified in our analysis. The gender bias in the lecture course coming from both men and women, therefore, is related to who gets recognized (particularly, students nominating their same-gender interaction ties) and not what gets recognized.

%prompting future work that more closely investigates how students develop perceptions of strong physics peers, such as through interviews.

\subsubsection{Gender bias in the lab course is related to both who and what gets recognized}

We found that the gender bias in the lab course (in which men disproportionately under-nominated women as strong in the course) is related to a gender bias in interaction-based recognition in the fall offering (similar to the patterns observed in the lecture courses described above), but is likely related to a gender bias in observation-based recognition in the spring offering (though we could not measure this). Different from the lecture course, the nature of peer interactions \textit{are also} a possible source of the gender bias in the lab course: men nominated men more than women for skills related to their interactions, such as helping them with the course material and explaining course material to others more generally. This result is consistent with prior work proposing that the ``chilly climate" for women in physics may be due to the nature of their peer interactions rather than their number of peer interactions~\cite{pearson2017developing}. It is surprising, however, that the men and women in our study proportionally nominated one another for specific experimental skills in lab, such as handling the equipment to collect data or leading the group, despite evidence of students' gendered engagement in these roles~\cite{danielsson2012exploring,gonsalves2016masculinities,Quinn2020}. 

%Different from the lecture course, we did not observe a gender bias from women in the lab course (i.e., women proportionately nominated men and women in both offerings of this course). In the spring offering, we did observe that women made disproportionately more observation-based nominations to men than women despite there not being a significant gender bias from women in this course. These nominations, however, likely did not outweigh the strong gender homophily already present in the interaction network, where ties from women to women were more common than ties from women to men. 

One possible explanation for why we observed a gender bias (from men) in the in-person lab course but not the remote labs (Course A in Ref.~\cite{sundstrom2022perceptions}) is that these two labs had very different course structures, which ultimately shaped opportunities for peer interactions and observations. In the remote course, the lab was attached to the lecture course and almost all work for lab was done during the lab sessions within lab groups in isolated breakout rooms on Zoom. The lab had no whole-class lecture session or summative assessments (e.g., exams or quizzes) and students' lab grades were combined with their lecture course grade. The in-person lab course, in contrast, included a whole-class lecture session that included peer instruction activities and three summative quizzes. Students also received a separate course grade for the lab material. We suspect that these changes in course structure facilitated more out-of-lab-group interactions, including both in-lecture interactions and out-of-class interactions. Out-of-class interactions may also have been more common during the in-person course simply due to the evolving nature of the COVID-19 pandemic.

\subsubsection{Summary}

The results from both courses add nuance to our previous work which found that the presence or absence of a gender bias in peer recognition varies between the instructional contexts of lab and lecture~\cite{sundstrom2022perceptions}. In this study, we observed a gender bias in peer recognition (either from both men and women or just men) in \textit{both} the lab and lecture course but found that the sources of this bias varied -- whether only due to a bias in who gets recognized (lecture) or also due to a bias in what gets recognized (lab). These results may suggest that pedagogical style (e.g., having both lecture and small-group sessions related to instructional material) may impact gender bias in peer recognition more than the instructional material itself. Future investigations of peer recognition, however, should continue to analyze peer recognition in these instructional contexts separately and should examine the nature of gender bias in peer recognition in other lab and lecture contexts, such as those at other institutions, with students from other types of majors, or with students in studio physics courses. Researchers should also more closely investigate how students develop perceptions of strong physics peers, such as through interviews, that may better illuminate some of the differences we observed between courses and semesters and between men's and women's nominations.

\subsection{Other implications for research}

Here we synthesize our findings that are not directly tied to the research questions of this study with those of previous research and suggest directions for future work.

\subsubsection{Structure of peer recognition networks}

The structures of the peer recognition networks in this study differ from those observed in prior work. Specifically, a relatively small fraction of nodes in the recognition networks of the in-person lecture courses in this study comprised the \textit{giant component} (the largest cluster of connected nodes), with connections instead forming short chain-like structures and some small, disconnected components. In previous studies~\cite{grunspan2016,salehi2019,sundstrom2022perceptions}, in contrast, the lecture recognition networks for both in-person and remote science courses had relatively large giant components (i.e., a large proportion of the nodes connected together in a main cluster). We propose that this difference in network structure is due to differences in course structure and populations. Refs.~\cite{grunspan2016,salehi2019}, for example, examined second-semester (or later) in-person courses within a multi-course sequence, when students have likely developed familiarity with one another. Thus, students could likely identify many of their peers, including outspoken students in lecture who then become celebrities in the network. Similarly, in the remote physics courses studied in Ref.~\cite{sundstrom2022perceptions}, Zoom likely expedited name familiarity and connections across many different peers (not just those in close physical proximity) even for students in the first course of the sequence. The in-person lecture courses analyzed in this study, however, are the first in the physics sequence when students are likely still getting to know one another and each other’s names (e.g., even if some students are really outspoken during in-person lectures, other peers likely do not know their name). Future work should examine this effect of peer familiarity on peer recognition further through a longitudinal study, for example modeling how recognition networks change (or not) throughout a course sequence (e.g., similar to Ref.~\cite{bruun2014time} in which the authors analyze changes in peer interaction networks over time).

We also observed lower transitivity (small group clustering) in the peer recognition networks of in-person lab courses (this study) than the peer recognition networks of remote physics labs~\cite{sundstrom2022perceptions}. This difference is likely due to the different course structures mentioned above and the different instructional modalities. In the remote physics labs~\cite{sundstrom2022perceptions}, students only attended low-enrollment lab sessions where they worked in Zoom breakout rooms with the same peers every week. Therefore, students had limited access to peers (and these peers' names) outside of their immediate lab group. Students in the in-person lab course (this study), however, could see and interact with other students in their lab section, including those outside of their immediate lab group. These students also attended a large lecture each week where they could interact with peers outside of their lab group and lab section entirely. Increased visibility of out-of-lab-group peers within individual lab sections and the addition of a lecture session to the lab, therefore, likely increased the number of peers students had access to which in turn reduced the amount of small, isolated clusters in the recognition network. Future work should investigate peer recognition in other course structures containing structured small group activities, such as studio physics and modeling instruction, where patterns of recognition might also vary.

\subsubsection{How does peer recognition form?}

We found that both peer interactions and indirect observations of peers play an important role in shaping peer recognition, in line with what prior work suggests~\cite{gee2000chapter,quan2022trajectory,alaee2022,grunspan2016}. Our analysis, therefore, illuminates the complexities behind how students form perceptions of their peers (i.e., interaction-based versus observation-based recognition). Future work should further investigate the relationship between interactions and peer recognition, for example by conducting egocentric network analyses to probe the more fine-grained social processes underlying the development of peer recognition. Our results also suggest that future analyses of peer recognition should continue to probe and analyze interaction networks alongside student nominations of strong peers. Examining both networks together will likely provide a more nuanced understanding of how patterns of recognition form.

Finally, we devised a coding scheme to identify the different skill sets for which students nominate their peers as strong in their physics course. These codes encompassed a variety of skills students identified in their explanations of nominations: knowledge (e.g., content knowledge, getting high grades, and having a natural ability to learn physics), engagement in processes (e.g., designing an experiment, analyzing data, and solving problems), and interactions (e.g., helping, explaining, and leading). Such skills have been identified across qualitative research studies investigating how students define being a good physics student or a good physicist~\cite{tonso2006student,danielsson2012exploring,fields2013picking,due2014competent,gonsalves2014physics,gonsalves2014persistent,gonsalves2016masculinities,irvingsayre2015,DoucetteHermione,doucettegoodlabpartner,cooper2018perceives,stump2022perc}, however our study is the first to evaluate these skills at the scale of a large introductory physics course. Future work should examine whether this coding scheme is applicable to explanations from students in other institutions or experiencing different instructional styles and pedagogies.

\subsection{Limitations}

We end this section by acknowledging the limitations of our study. First, the online network survey may not have captured all nominations of strong peers and all peer interactions. Students may not have remembered the names of individuals they perceived as strong in the material and/or with whom they interacted, for example due to recall bias. We also only collected survey responses in the middle of the semester. Previous work administered surveys either both at the middle and end of the course~\cite{grunspan2016} or only at the end of the course~\cite{salehi2019} and so this methodological choice allowed us to compare our results to that previous work. Future research examining in-person physics students' recognition of strong peers at multiple points in their physics course, or just at the end of their physics course, however, may add nuance to our results.

Additionally, we categorized students' race or ethnicity by URM status because the number of students in some of the individual racial or ethnic groups was too small for our statistical analysis to produce useful and interpretable results. This treatment of race or ethnicity, however, inevitably masks differences in peer recognition between students of individual racial or ethnic groups. Because our research questions were focused on the role of gender in peer recognition, this categorization allowed us to account for race and ethnicity in a cursory way in our analysis. Future work should seek to study the role of race and ethnicity in peer recognition explicitly by using more diverse student populations with statistically sufficient representation across racial or ethnic groups. Researchers should also aim to differentiate peer recognition among White and Asian students and between Asian and Asian American students~\cite{wing2007beyond,cabrera2014beyond}. 

We also observed more sparse recognition networks than those in prior work~\cite{grunspan2016,salehi2019,sundstrom2022perceptions}, particularly with regard to the proportion of students who were isolates (nodes with zero adjacent edges in the network). It is impossible for us as researchers to know if these isolates are ``true isolates," students who truly have zero nominations to make or receive, or if they are essentially non-respondents who fill out the survey quickly and do not nominate anyone, even though they may actually have nominations to make. Therefore, our analysis may have missed some nominations and the network structures may not represent all connections between students. Fortunately, social network data is robust to up to 30\% missing data~\cite{smith2013structural} and our recognition networks contain a range of 32\%--48\% isolates. Assuming some students are true isolates, our analysis likely captures a fairly accurate picture of peer recognition. Furthermore, all of our ERGMs converged with appropriate goodness-of-fit diagnostics, indicating that the statistical analysis is reliable; any uncertainties due to small sample sizes are reflected in the standard errors and $p$-values of the coefficient estimates.

Finally, the majority of our analysis assumes that any overlap in the recognition and interaction networks implies that peer recognition was formed through peer interactions. Alternatively, interactions may plausibly be formed through recognition, such as by a student deciding to interact with peers they perceive as strong in the course material. We believe the former is more likely than the latter in this study because the courses we analyze are the first in the course sequence, when students have likely not met each other before, and because the survey was administered in the middle of the semester, when students may not have a strong sense of each other's grades. Nonetheless, future work should seek to disentangle this relationship between interactions and recognition, if possible.

\section{Conclusion}

In this study, we observed a gender bias in student nominations of strong physics peers in distinct lab and lecture courses. In both courses, men under-nominated women as strong in the course. Women also under-nominated men as strong in one offering of the lecture course. To understand the nature of this gender bias in peer recognition, we built upon two existing threads of research. First, we investigated the role of peer interactions in students' determination of who gets recognized in peer recognition. Second, we examined what gets recognized -- whether the skills students associated with being a strong physics student or strong physicist are related to the gender bias in peer recognition. We found that roughly half of nominations of strong peers formed through peer interactions, with the remaining nominations likely coming from indirect observations of peers. We also observed that the nature of the gender bias varied between the lab and lecture courses. In the lecture course, the bias was related to who gets recognized: both men and women disproportionately over-nominated their interaction ties with students of their same gender as strong in the course material. At the same time, men and women nominated men and women for similar skill sets in this course. In the lab course, in contrast, men also disproportionately under-nominated women for certain skill sets, particularly those related to their interactions, such as being helpful. These results add nuance to our understanding of how students form perceptions of strong peers in their physics courses and prompt future work to examine the nature of gender bias in peer recognition in other course structures, instructional styles, and student populations.

\section*{ACKNOWLEDGEMENTS}

This material is based upon work supported by the National Science Foundation Graduate Research Fellowship Program Grant No. DGE-2139899 and Grant No. DUE-1836617. We thank Matthew Dew and David Esparza for providing meaningful feedback on this work.

\bibliography{perceptionsexplanations.bib} 

\section*{Appendix}

\subsection{ERGMs: Goodness-of-fit diagnostics}

\begin{figure}[t]
    \centering
    \includegraphics[trim={0 0 16cm 0},scale=0.39]{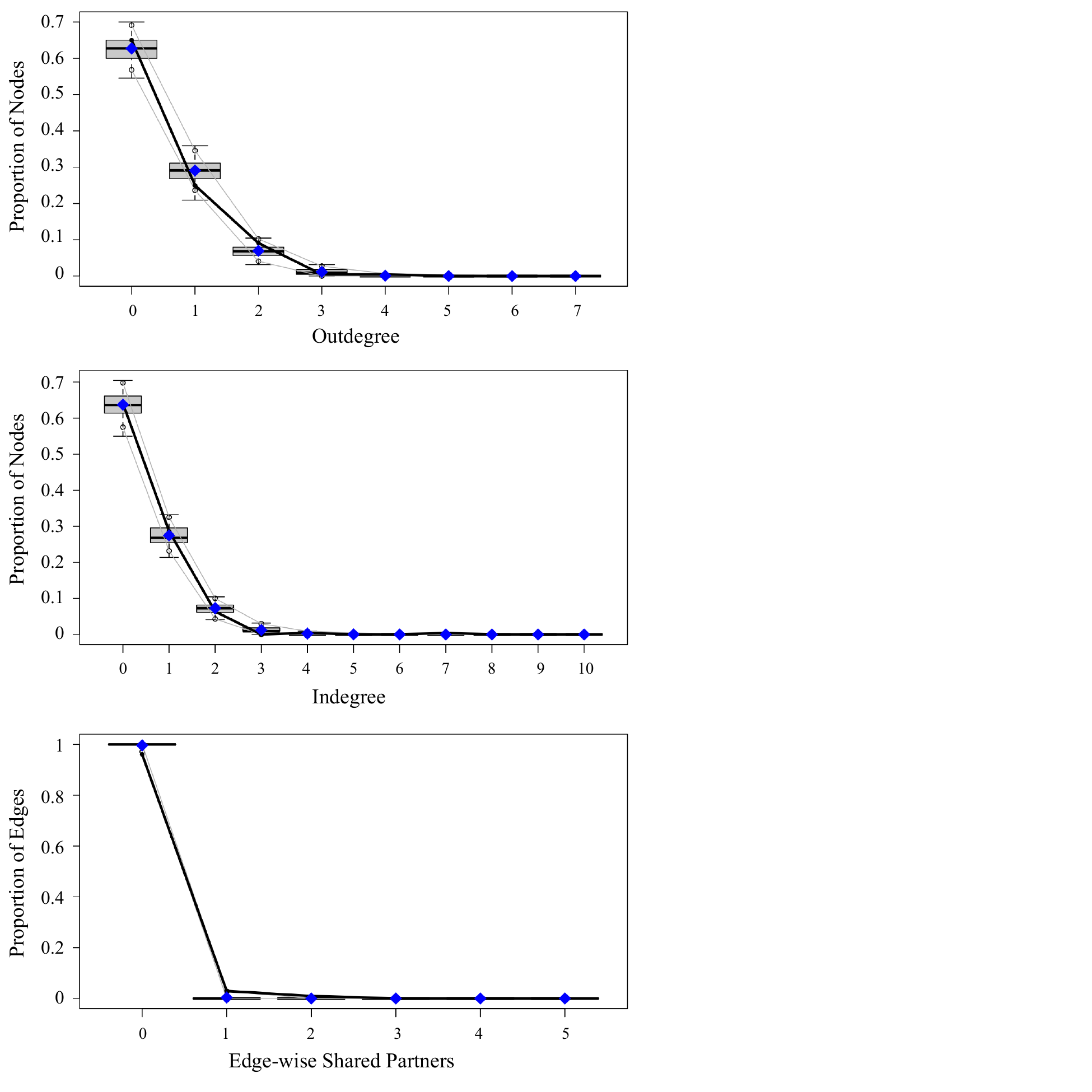}
    \caption{ERGM goodness-of-fit diagnostics for the recognition network of the fall offering of the lecture course. The horizontal axis represents a network measure (outdegree, indegree, or edge-wise shared partners -- a measure of transitivity) and the vertical axis represents either the proportion of nodes or proportion of edges in the network. Plots compare the distribution of each network measure for the observed data (thick black line) to that for 10 network simulations generated using the estimated model coefficients (boxplots).}
    \label{fig:gof}
\end{figure}

Figure~\ref{fig:gof} shows one example (for the fall offering of the lecture course) of the diagnostics used to assess the goodness-of-fit of ERGMs. When an ERGM fits an observed network well, the thick black line always falls within the boxplots. Because this is the case in Fig.~\ref{fig:gof} and for the other three analyzed courses, our model sufficiently captures the characteristics of the observed networks.

\subsection{ERGMs: Full model results}

Table~\ref{tab:coefvalues} shows the coefficient estimates for all predictor variables in our exponential random graph models, fit to each of the four analyzed networks.

\begin{table*}[t] 
\centering
\caption{\label{tab:coefvalues} Coefficient estimates for the exponential random graph models, represented as log-odds, fit to each observed recognition network. Standard errors are given in parentheses. Asterisks indicate statistical significance ($^{*} p < $0.05; $^{**} p < $0.01).}
\begin{ruledtabular}
\setlength{\extrarowheight}{3pt}
\begin{tabular}{lcccc}
 & \multicolumn{2}{c}{Lab course} &  \multicolumn{2}{c}{Lecture course}  \\ 
 \cline{2-3}
 \cline{4-5}
 & Fall  & Spring  & Fall  & Spring  \\ 
 \hline
 \textit{Edges} & --10.37$^{**}$ (0.73) & --11.11$^{**}$ (0.56) & --9.49$^{**}$ (0.87) & --10.70$^{**}$ (0.48) \\ 
\textit{Reciprocity} & --0.14 (0.31) & 0.46$^{*}$ (0.23) & 2.95$^{**}$ (0.52) & 4.84$^{**}$ (0.32)  \\ 
\textit{Woman $\rightarrow$ woman} & --0.04 (0.19) & 0.04 (0.13) & 0.19 (0.22) & 0.15 (0.15)  \\ 
\textit{Woman $\rightarrow$ man} & 0.02 (0.19) & 0.02 (0.15) & --0.57 (0.30) & --0.41$^{*}$ (0.20)  \\ 
\textit{Man $\rightarrow$ woman} & --0.83$^{**}$ (0.25) & --0.42$^{*}$ (0.17) & --0.98$^{**}$ (0.36) & --1.06$^{**}$ (0.26)  \\ 
\textit{URM $\rightarrow$ URM} & 0.19 (0.44) & --0.23 (0.24) & 0.92$^{*}$ (0.41) & 0.47 (0.25)  \\ 
\textit{URM $\rightarrow$ non-URM} & 0.01 (0.25) & 0.10 (0.15) & 0.19 (0.29) & --0.17 (0.20)  \\ 
\textit{Non-URM $\rightarrow$ URM} & --0.25 (0.26) & 0.08 (0.16) & 0.30 (0.31) & --0.11 (0.24)  \\ 
\textit{Physics majors $\rightarrow$ physics majors} & 1.36$^{**}$ (0.41) & 1.59$^{**}$ (0.43) & N/A & N/A  \\ 
\textit{Physics majors $\rightarrow$ non-majors} & --0.96$^{*}$ (0.40) & --0.07 (0.28) & N/A & N/A  \\ 
\textit{Non-majors $\rightarrow$ physics majors} & 0.20 (0.25) & 0.69$^{**}$ (0.21) & N/A & N/A  \\
\textit{Lab group homophily} & 4.64$^{**}$ (0.26) & 4.20$^{**}$ (0.18) & 1.70$^{**}$ (0.56) & 2.07$^{**}$ (0.38)  \\ 
\textit{Lab section homophily} & 1.96$^{**}$ (0.27) & 2.50$^{**}$ (0.20) & --0.04 (0.47) & 0.17 (0.33)  \\ 
\textit{Discussion section homophily} & N/A & N/A & 2.14$^{**}$ (0.21) & 1.62$^{**}$ (0.32)  \\ 
\textit{Grade of nominee} & 0.71$^{**}$ (0.19) & 0.75$^{**}$ (0.15) & 0.79$^{**}$ (0.23) & 1.01$^{**}$ (0.13)  \\  
\end{tabular} 
\end{ruledtabular}
\end{table*}

\subsection{Interaction network diagrams}

\begin{figure*}[t]
    \centering
    \includegraphics[width=6.6in]{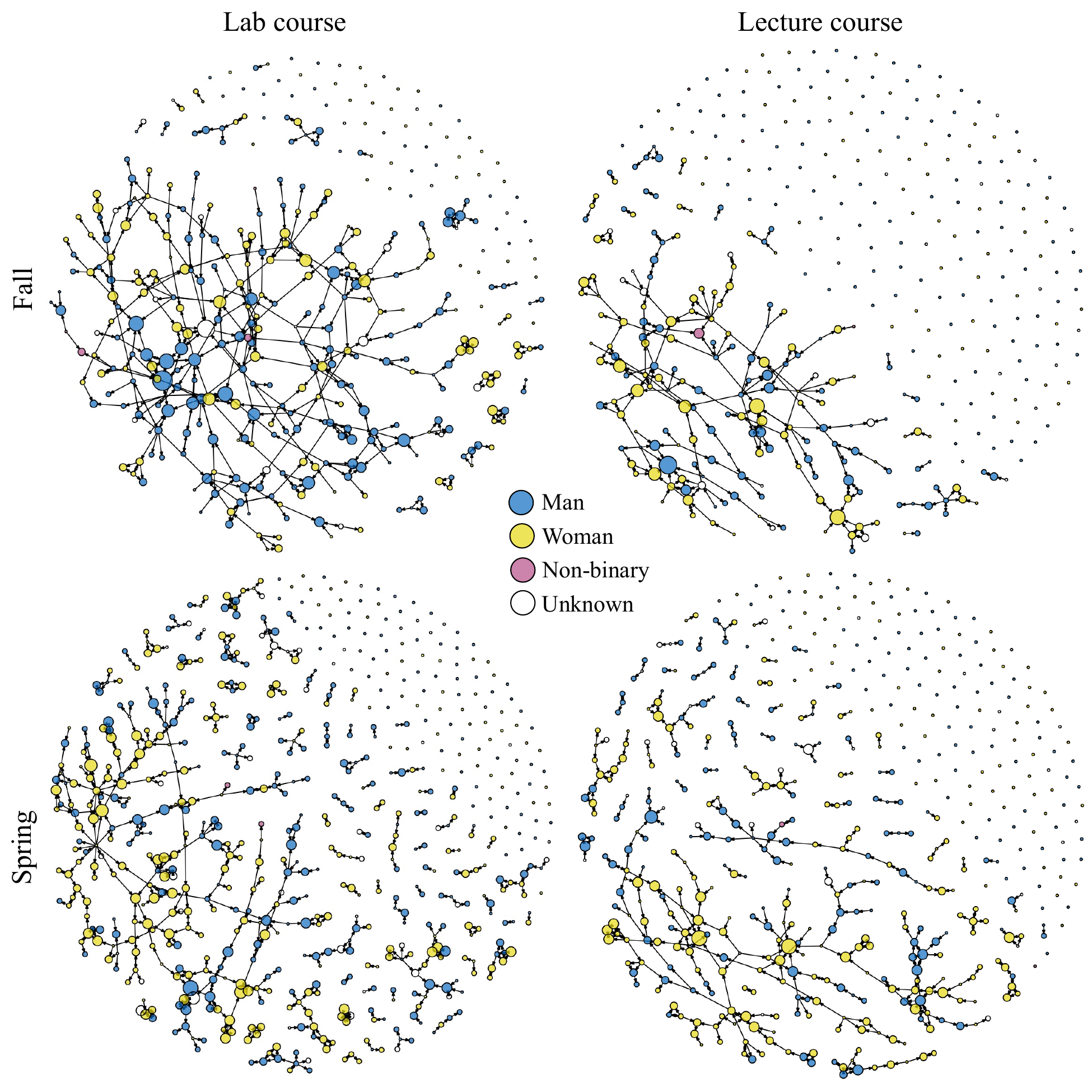}
    \caption{Interaction networks for all analyzed courses. Nodes are colored by gender and sized proportional to indegree (number of received nominations). Edges point from the nominator to the nominee.}
    \label{fig:genderintsociograms}
\end{figure*}

Figure~\ref{fig:genderintsociograms} shows the network diagrams of students' self-reported peer interactions for all four analyzed courses.

\subsection{Explanations coding scheme: \textit{Other} code examples}

We coded an explanation as \textit{other} if it was not related to one of the other codes (see Table~\ref{tab:codingscheme}) or if the reason(s) provided in the explanations did not appear often enough in the full data set to form a new code. Examples of vague explanations coded as \textit{other} are ``Good at working in the lab'' and ``She's amazing." Examples of ideas that were too infrequent to form their own code are:
\begin{itemize}
    \item drawing (e.g., ``She draws good.")
    \item open-minded (e.g., ``[He] is open-minded and explains concepts materials when I don't understand.")
    \item patient (e.g., ``Patient with other members.")
    \item listening (e.g., ``He listens to new ideas and knows the course material very well.")
    \item revising their own thinking (e.g., ``She understands and is confident in the course material and can iteratively revise her thinking.")
    \item creative (e.g., ``He quickly comes up with good ideas for experiments and is very creative with collecting data.")
    \item hearing from someone else that the nominee is strong in the course (e.g., ``I’ve heard great things of this man.").
\end{itemize}

%\subsection{Code frequencies in lecture course by gender}

%\begin{figure}[t]
 %   \centering
 %   \includegraphics[width=3.3in,trim={1cm 0 15cm 0}]{GenderLectureCodes.pdf}
 %   \caption{Fraction of nominations falling under each code of our coding scheme within the lecture course, split by gender of the nominator and nominee, split by gender of the nominator and nominee.}
 %   \label{fig:genderlecturecode}
%\end{figure}

\subsection{Explanations split by interaction-based and observation-based nominations}

\begin{figure*}[t]
    \centering
    \includegraphics[width=6.7in,trim={0 0 2cm 0}]{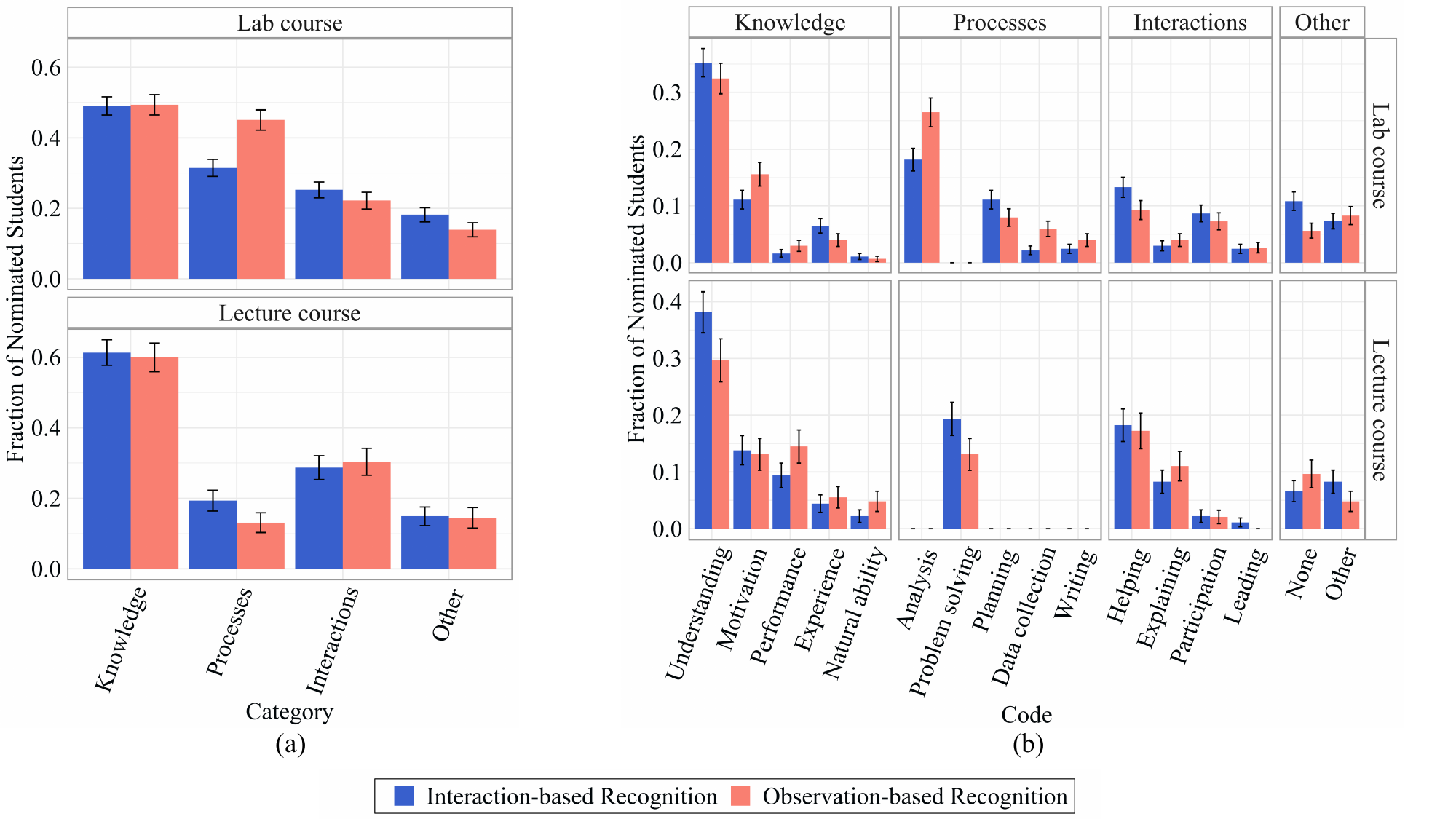}
    \caption{Fraction of nominations that are interaction-based and observation-based falling under each (a) category and (b) code of our coding scheme, split by course.  Fractions do not necessarily sum to one because each explanation could receive multiple codes. Error bars represent standard errors of the proportions. }
    \label{fig:explanationsappendix}
\end{figure*}

Figure~\ref{fig:explanationsappendix} shows the fraction of interaction-based nominations (i.e., nominations of strong peers with whom the nominee also reported interacting) and observation-based nominations (i.e., nominations of strong peers with whom the nominee did not report interacting) falling under each category or code from our coding scheme. 

\end{document}